\documentclass[preprint,twocolumn,astrosymb]{aastex631}

\usepackage{macros}

\usepackage{amsmath}
\usepackage{amsfonts}
\usepackage{amssymb}
\usepackage[caption=false]{subfig}
\usepackage{graphicx}
\usepackage{multirow}

\definecolor{midgreen}{rgb}{0.0,0.675,0.0}

\graphicspath{{./}{figures/}}
\shortauthors{Farah, Fishbach \& Holz}

\begin{document}

\title{Two of a Kind: Comparing big and small black holes in binaries with gravitational waves}

\author[0000-0002-6121-0285]{Amanda M. Farah}
\email{afarah@uchicago.edu}
\affiliation{Department of Physics, University of Chicago, Chicago, IL 60637, USA}

\author[0000-0002-1980-5293]{Maya Fishbach}
\affiliation{Canadian Institute for Theoretical Astrophysics, David A. Dunlap Department of
Astronomy and Astrophysics, and Department of Physics, 60 St George St, University of Toronto, Toronto, ON M5S 3H8, Canada}

\author[0000-0002-0175-5064]{Daniel E. Holz}
\affiliation{Department of Physics, University of Chicago, Chicago, IL 60637, USA}
\affiliation{Kavli Institute for Cosmological Physics, The University of Chicago, 5640 South Ellis Avenue, Chicago, Illinois 60637, USA}
\affiliation{Enrico Fermi Institute, The University of Chicago, 933 East 56th Street, Chicago, Illinois 60637, USA}

\begin{abstract}
When modeling the population of merging binary black holes, analyses have generally focused on characterizing the distribution of primary (i.e. more massive) black holes in the binary, while simplistic prescriptions are used for the distribution of secondary masses.
However, the secondary mass distribution \new{and its relationship to the primary mass distribution} provide a fundamental observational constraint on the formation history of coalescing binary black holes.
If both black holes experience similar stellar evolutionary processes prior to collapse, as might be expected in dynamical formation channels, the primary and secondary mass distributions would show similar features.
If they follow distinct evolutionary pathways (for example, due to binary interactions that break symmetry between the initially more massive and less massive star), their mass distributions may differ.
\new{We present the first analysis of the binary black hole population that explicitly fits for the secondary mass distribution.
We find that the data is consistent with a $\sim30\,M_{\odot}$ peak existing only in the distribution of \emph{secondary} rather than primary masses.
This would have major implications for our understanding of the formation of these binaries.
Alternatively, the data is consistent with the peak existing in both component mass distributions, a possibility not included in most other previous studies.
In either case, the peak is observed at $31.4_{-2.6}^{+2.3}\,M_{\odot}$, which is shifted lower than the value obtained in previous analyses of the marginal primary mass distribution}, placing this feature in further tension with expectations from a pulsational pair-instability supernova pileup.
\end{abstract}


\section{Introduction} \label{sec:intro}
Dozens of gravitational-wave (GW) events have been observed by the LIGO \citep{aasi_advanced_2015}, Virgo \citep{acernese_advanced_2014} and KAGRA~\citep{akutsu_overview_2021} detector network \citep{abbott_gwtc-3_2021}, and many more detections are anticipated by the end of the fourth observing run.
However, the formation mechanism of binary black hole (BBH) mergers, which source the majority of detected GW events, is still largely uncertain.
While it is not possible to know the formation history of any one BBH with certainty, the population of all merging BBHs encodes information about which astrophysical processes give rise to the bulk of these systems \citep[e.g.][]{stevenson_distinguishing_2015, zevin_constraining_2017}.

The stellar-mass BBHs detectable by LIGO, Virgo and KAGRA are likely created from the collapse of massive stars. 
These massive stars may be born as a binary system in the galactic field that then evolves into a BBH system.
Alternatively, they may be born in a dense stellar environment, such as a star cluster, in which the BH stellar remnants dynamically assemble into tightly-bound binaries.
The population of GW sources contains signatures of the initial conditions of their progenitor stellar systems, as well as many of the evolutionary processes that occur between star formation and BBH merger.
The initial conditions that impact the GW source population include the initial mass function of either binary or single stars (binary IMF and IMF, respectively) in different environments, the birth metallicities of the progenitor stars, and other aspects of their formation environments.
Depending on the BBH formation scenario, the evolutionary processes that are imprinted on the BBH population include stellar mass loss, transfer of matter between two component stars, the supernova process for massive stars, and dynamical interactions in star clusters or the disks of active galactic nuclei \citep[see reviews by][and references therein]{mapelli_binary_2020,mandel_merging_2022}.

In general, these uncertain formation processes affect the masses, spins, redshifts, eccentricities, and merger rates of BBH systems. 
Here we focus on the mass distribution, since BH masses are well-measured from the GW signal.

The BBH mass distribution is typically parameterized by the primary mass $m_1$, the larger of the two component masses in the binary, and either the secondary mass $m_2$, or the mass ratio $q=m_2/m_1$.
Within these parametrizations, it is typically assumed either that the primary and secondary masses follow the same underlying distribution \citep{fishbach_picky_2020,doctor_black_2020,farah_bridging_2022,edelman_cover_2022,abbott_population_2023, sadiq_binary_2023}, or that the primary mass distribution has distinct features from the secondary mass distribution.
The latter assumption implies that primary and secondary mass are physically meaningful labels, and is typically achieved by modeling the secondary mass distribution as a single power law between some minimum mass and $m_1$ \citep{fishbach_where_2017,kovetz_black_2017,talbot_measuring_2018,baxter_find_2021,abbott_population_2021,tiwari_vamana_2021,abbott_population_2023, edelman_aint_2022,callister_parameter-free_2023,godfrey_cosmic_2023}. 
The former assumption---that $m_1$ and $m_2$ follow the same underlying distribution---implies that the two-dimensional mass distribution is symmetric in primary and secondary mass, meaning that $m_1$ and $m_2$ can be interchanged without changing the mass distribution.
Neither of these assumptions have been explicitly verified with the available data. 
In this work, we aim to determine if both components in merging BBHs follow the same underlying distribution or if there is a physical distinction between primary and secondary masses.

Understanding whether the primary and secondary masses in BBH mergers follow the same distribution will provide insight into their formation histories.
For BBH mergers that are dynamically assembled in dense environments, we expect that both component BHs are drawn from the same population of stellar remnants (i.e. $m_1$ and $m_2$ are not physically meaningful labels), and that the two-dimensional BBH mass distribution will therefore be symmetric in $m_1$ and $m_2$.
For BBH mergers that originate from binary stars that formed and evolved in relative isolation (``field binaries"), there may be a physical distinction between primary and secondary BH masses. 
To start off, the progenitor stars in the binary are drawn from the binary IMF, which may not be symmetric between the two components \citep[e.g.][]{grudic_does_2023}. 
Furthermore, the two stars exchange mass during binary stellar evolution. 
In each phase of mass transfer, one component acts as the donor and the other as the accretor, depending on their initial masses. 
Mass transfer affects the donor and the accretor stars in different ways, which can impact the masses of the resulting BHs following stellar collapse \citep{laplace_different_2021}.
\new{On the other hand, both BH progenitors are expected to have undergone binary stripping and therefore experience similar supernova physics, potentially washing out any significant differences between the mass distributions of first- and second-born BHs \citep{van_son_redshift_2022, schneider_bimodal_2023}.
However, the degree to which these distributions are similar or different depends on uncertain mass loss and accretion physics \citep{ van_son_no_2022}.}

\new{Supernova kicks may also be different for first versus second-born components in \emph{merging} binaries, because kicks determine whether or not the binary can merge within the age of the Universe \citep{kalogera_orbital_1996, gallegos-garcia_evolutionary_2022}.
Because natal kicks are related to the remnant mass via the supernova prescription~\citep{fryer_compact_2012, mandel_binary_2021}, different preferences for the natal kick magnitudes between first-born and second-born BHs may cause the primary and secondary mass distributions to differ in merging binaries formed through isolated binary evolution \citep[e.g.][]{oh_role_2023}.}

In short, binary stellar evolution consists of several processes that can break the symmetry between the population of initially more massive stars, which generally correspond to the first born and more massive (primary) BHs, and the population of initially less massive stars, which generally correspond to the second born and less massive (secondary) BHs. 
However, if mass inversion occurs in some systems, some initially less massive stars will end up as the more massive BH by the time of merger, and the distribution of primary BH masses will have contributions from both the second-born and first-born BHs ~\citep{olejak_implications_2021,hu_channel_2022, broekgaarden_signatures_2022, zevin_suspicious_2022}. 
If mass inversion happens in exactly half of merging BBH systems, the primary and secondary component mass distributions may be indistinguishable even if the first and second born distributions differ.

From a data analysis perspective, knowing that the primary and secondary mass distributions are the same allows us to measure a single set of model parameters: those of the shared distribution.
This may allow features of the distribution to be measured with higher precision since both components in the binary will contribute to the measurement of each feature, rather than just the primary mass.
Furthermore, disentangling the role of primary and secondary masses aids the physical interpretations of such features.

For example, \citet{abbott_population_2023} found that the mass distribution exhibits a peak at primary masses $m_1\sim35\,M_\odot$.
The astrophysical origin of this overdensity in the mass distribution is uncertain, although it may be related to (pulsational) pair-instability supernovae~\citep[e.g.][]{heger_nucleosynthetic_2002,fishbach_where_2017-1,talbot_measuring_2018,farmer_mind_2019}.
A necessary ingredient towards understanding the origin of the $m_1\sim35\,M_\odot$ peak is to first understand whether the peak appears exclusively among primary BBH masses, or whether secondary masses also display a peak, indicating that secondary BHs also experience the astrophysical process that leads to a mass pileup.

This paper is organized as follows. 
In Section~\ref{sec:methods} we describe the different population models considered in this work.
In Section~\ref{sec:results} we present the results of fitting each of the models to the Third Gravitational-Wave Transient Catalog \citep[GWTC-3][]{abbott_gwtc-3_2021}, finding that while the primary and secondary masses of merging BBHs are consistent with following the same underlying distribution, it is also possible that the secondary mass distribution has a more prominent peak than does the primary mass distribution.
In Section~\ref{sec:future} we discuss possibilities for future observations.
In Section~\ref{sec:discussion} we summarize our conclusions and present possible astrophysical interpretations of our results.
The appendices discuss the fundamental differences between commonly-used parametrizations for the two-dimensional mass distribution and include details about the population models and statistical framework.

\section{Population Models} \label{sec:methods}
Because our aim is to learn whether the primary and secondary masses are consistent with being drawn from the same distribution, we describe the primary and secondary mass distributions separately, but according to the same functional form.
We model each of the one-dimensional mass distributions as a mixture model between a smoothed power law component and a Gaussian component in order to make direct comparisons to the \textsc{Power Law + Peak} model used by the LIGO-Virgo-KAGRA collaboration (LVK) to describe the distribution of primary masses \citep{talbot_measuring_2018,abbott_population_2021,abbott_population_2023}.

There are several ways to construct a two-dimensional mass model for the binary from this one-dimensional mass model for the component masses.
However, as we discuss in Appendix~\ref{ap:conditioned q}, in order to explicitly compare primary and secondary mass distributions, it is necessary to use the pairing function framework first introduced for GW population modeling in \citet{fishbach_picky_2020}.
Explicitly,
\begin{equation}
    p(m_1,m_2|\Lambda) = p_1(m_1|\Lambda_1)p_2(m_2|\Lambda_2)f(q;\beta_q),
\end{equation} 
where $p_1(m_1|\Lambda_1)$ is the underlying distribution of primary masses, $p_2(m_2|\Lambda_2)$ is the underlying distribution of secondary masses, $f(q)$ is a pairing function that depends on the mass ratio of the system\footnote{In principle, the pairing function can be parameterized in terms of any observable parameter (e.g., total mass).
}, $\Lambda_1$ and $\Lambda_2$ are the hyper-parameters\footnote{We use the term ``hyper-parameter'' for a model parameter that describes the population of merging compact binaries, in contrast to the parameters describing each individual GW detection, such as one system's masses and spins.} describing the underlying primary and secondary mass distributions, respectively, and $\Lambda = \{\Lambda_1, \Lambda_2, \beta_q\}$ is the set of all mass model hyper-parameters.
In this work, we use a pairing function of the form $f(q;\beta_q) = q^{\beta_q} \Theta(q\leq1)$, though other forms may provide a better fit to the data \citep[e.g.][]{farah_bridging_2022}. 

\begin{figure*}
    \centering
    \includegraphics[width=\textwidth]{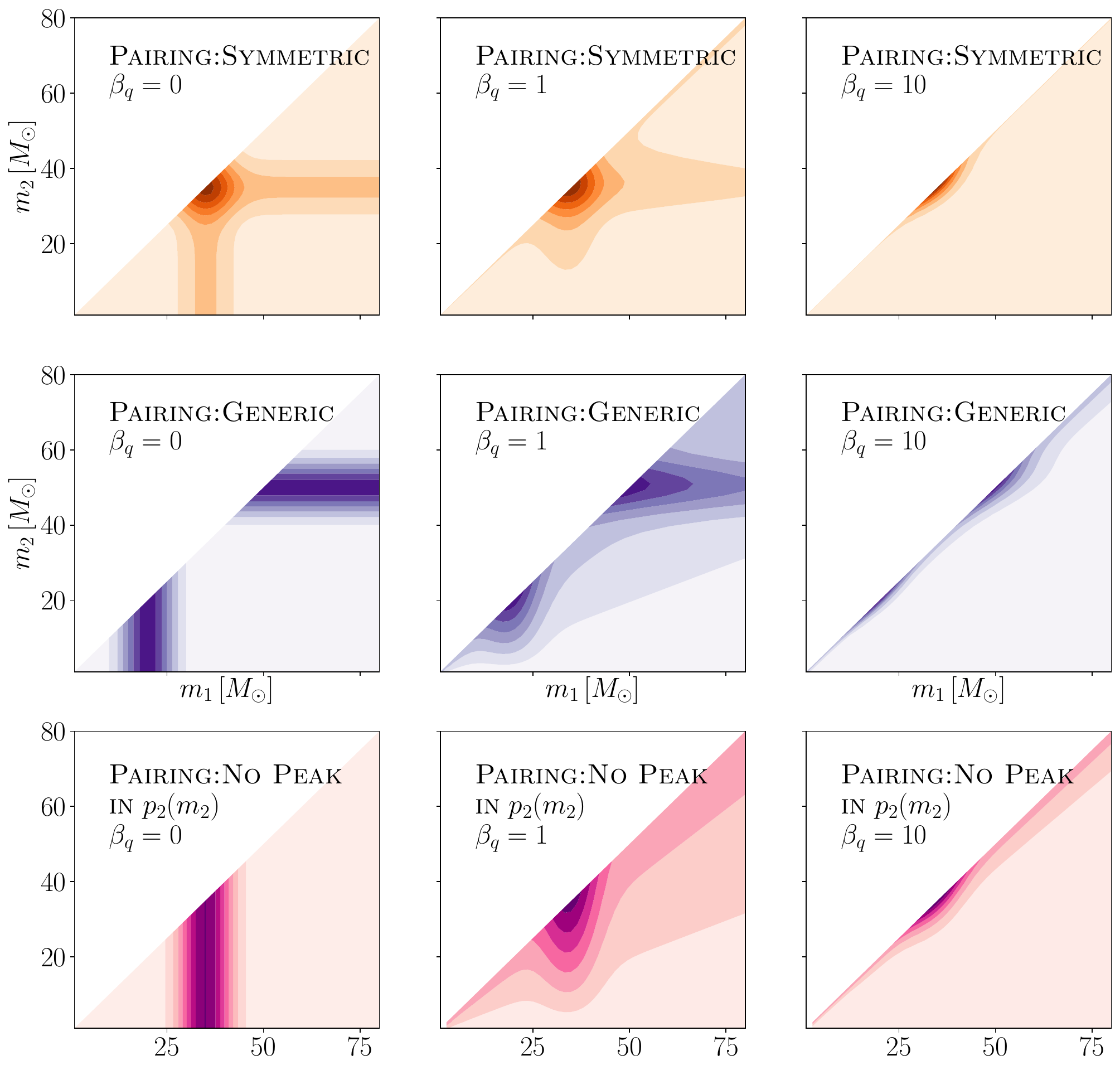}
    \caption{Illustration of some possible two-dimensional mass distributions under the models considered in this work.
    \new{These distributions are not indicative of specific astrophysical predictions, but are instead meant to be illustrative of the morphologies accessible by current models.}
    Overdensities/peaks in the mass distribution appear as darker filled contours in these figures.
    The different columns correspond to different power law spectral indices for the mass ratio-dependent pairing function, $\beta_q$.
    In the case where components strongly prefer to pair with nearly equal-mass partners, it becomes difficult to determine whether a feature appears only in one component mass distribution (as in the \lamTwoeqZero{} model, \emph{bottom row}) or in both (\default{}, \emph{top row}).
    The diagonal contours in the middle and right columns are caused by a preference for equal-mass binaries and follow lines of constant mass ratio.
    The goal of this paper is to distinguish between the different scenarios illustrated in this figure.
    }
    \label{fig:cartoon}
\end{figure*}

As illustrated in Figure~\ref{fig:cartoon}, different choices for the relationship between $\Lambda_1$ and $\Lambda_2$ result in distinct morphologies for the two-dimensional mass distribution. 
Below, we list each of the variations we consider in this work, along with the panels in Figure~\ref{fig:cartoon} to which they correspond.
A table describing these variations in terms of choices for the hyper-prior is given in Appendix~\ref{ap:model comparison}.
\begin{itemize}       
    \item \default: This model sets $\Lambda_1=\Lambda_2$, making the distribution symmetric under the transformation $m_1 \leftrightarrow m_2$.
    It corresponds to the first row of Figure~\ref{fig:cartoon}: any feature in one of the distributions has to be present in both, so it always makes two bands that connect on the diagonal.

    \item \diff: This model allows the hyper-parameters describing the $\sim35\Msun$ peak to differ between $p_1(m_1|\Lambda_1)$ and $p_2(m_2|\Lambda_2)$.
    It can produce any of the panels in Figure~\ref{fig:cartoon}, and is the only model that can produce a scenario such as that illustrated in the middle panel, where the feature in $m_1$ has a different amplitude and location than the feature in $m_2$. 

    \item \lamTwoeqZero{}: 
    This model sets all $\Lambda_2=\Lambda_1$, except for the parameter governing the height of the $\sim35\Msun$ peak, which we set to vanish for the secondary mass distribution but fit as a free parameter for the primary mass distribution.
    This model corresponds to the bottom row of Figure~\ref{fig:cartoon}: it is only capable of having a peak in the primary mass distribution, so can only produce vertical bands in a two-dimensional mass distribution.
    In Appendix~\ref{ap:all_same_lam2_0}, we show that \lamTwoeqZero{} approximately mimics the behavior of the commonly-used \textsc{Power Law + Peak} model from, e.g., \citet{abbott_population_2023}, which we refer to as \qdist. 
\end{itemize}

The different columns in Figure~\ref{fig:cartoon} correspond to different power law spectral indices, $\beta_q$, for the mass ratio-dependent pairing function.
The leftmost panels show models where components in the binary are allowed to pair up independently of mass ratio, the middle column shows a model where components have a slight preference to pair up with partners that are equal mass, and the rightmost panel shows the case where components are very ``picky'': they almost always pair up with equal mass partners \citep{fishbach_picky_2020}.
When components pair nearly independently of mass ratio, the asymmetric models produce noticeably different distributions than the symmetric models.
However, in the case of very picky binaries, the two scenarios become difficult to tell apart.
There is therefore a degeneracy between the steepness of the pairing function and the existence of distinct features in the two mass distributions \citep[see][for a discussion of this phenomenon in terms of Jacobian transformations]{tiwari_whats_2023}.
 
In all models considered in this work, the redshift distribution is modeled as a power law with spectral index $\kappa$ \citep{fishbach_does_2018}.
We use the \textsc{Default} spin model from \citet{abbott_population_2021,abbott_population_2023} to describe the spin magnitudes and tilts of each component.
These are the same redshift and spin distributions used with \qdist{} in \citet{abbott_population_2023}.
The explicit form of the full population model is given in Appendix~\ref{ap:most general model}.

Using these parameterized models, we construct a hierarchical Bayesian inference (described in Appendix~\ref{ap:HBA}) to determine the appropriate population-level parameters for the mass distribution, $\Lambda$, given the observed set of data $\{D_j\}$ for $N$ observed events \citep{loredo_accounting_2004, mandel_extracting_2019}.
We model the data as an inhomogeneous Poisson process with the rate density (number of events per unit time per single-event-parameter hypervolume) given by
\begin{equation}
    \frac{d\mathcal{N}}{dm_1 dm_2, ds_1 ds_2 dz} = \mathcal{R} p(z) p(s_1,s_2) p(m_1,m_2|\Lambda) ,
\end{equation}
where $\mathcal{R}$ acts as a normalizing constant that sets the overall magnitude of the rate.

\section{Results}
\label{sec:results}
\begin{figure*}
    \includegraphics[width=\textwidth]{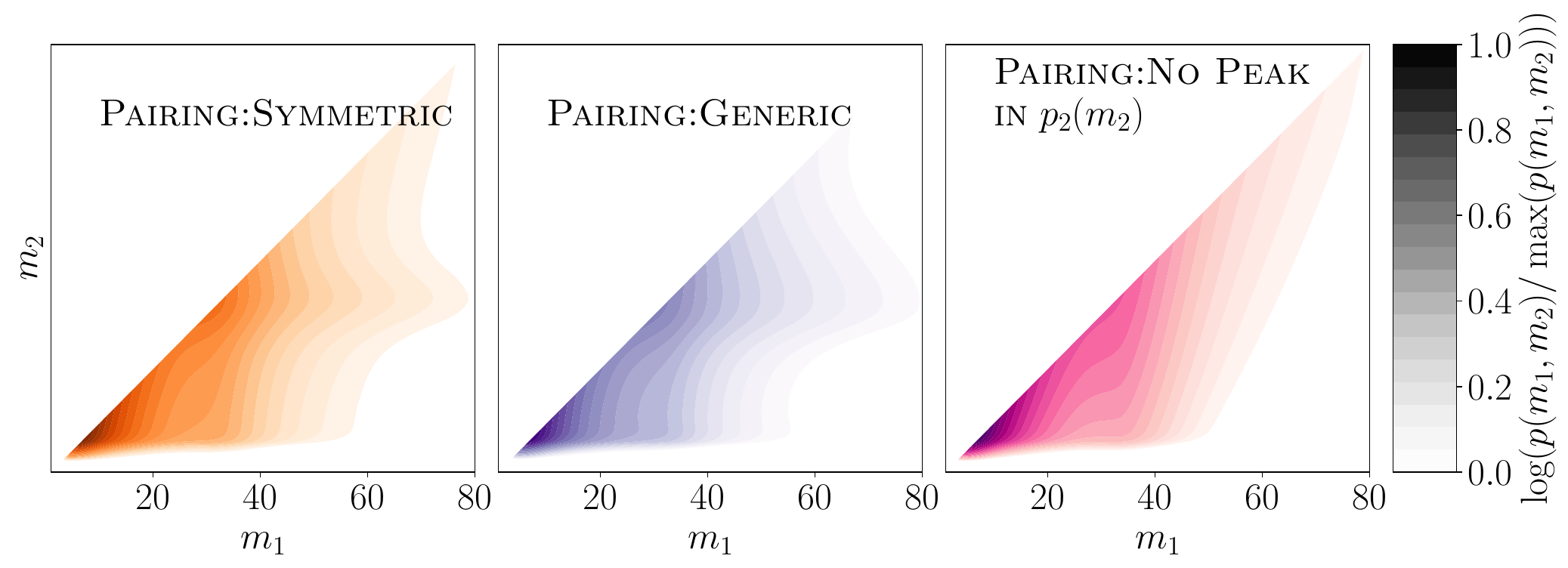}
    \caption{Two-dimensional posterior population distributions (PPDs) for \default{} (\emph{left, orange}), \diff{} (\emph{middle, purple}), and \lamTwoeqZero{} (\emph{right, pink}), all fit to the BBHs in GWTC-3.
    Darker colors indicate a higher rate of sources, and each panel is individually normalized to it maximum value.
    All three models find a higher rate of events near the $m_1=m_2$ diagonal, as well as peaks at $m_1\sim10\Msun$ and $m_1\sim35\Msun$.
    \default{} and \diff{} both find peaks in $m_2$ as well, as indicated by horizontal bands in the two \new{leftmost} panels, whereas \lamTwoeqZero{} is unable to model features in the $m_2$ direction.
    Additionally, \default{} and \diff{} display peaks in similar locations despite the fact that \default{} requires that both $m_1$ and $m_2$ follow the same underlying distribution but \diff{} is able to model each component separately.
    This suggests that both components may be drawn from the same underlying distribution, up to a pairing function. 
    }
    \label{fig:2d ppds}
\end{figure*}

We fit each model described in Section~\ref{sec:methods} to the BBHs in GWTC-3.
The resulting two-dimensional mass distributions for the \default{} and \diff{} models are shown in Figure~\ref{fig:2d ppds}.
These plots represent an average over the hyper-posterior for each model; this average is sometimes referred to as a posterior population distribution (PPD). 
The contours differ in morphology to those illustrated in Figure~\ref{fig:cartoon} because the actual distribution of BBH component masses exhibits two peaks: one at $\sim10\Msun$ and another at $\sim35\Msun$ \citep{abbott_population_2021, tiwari_vamana_2021,edelman_aint_2022,sadiq_flexible_2022,abbott_population_2023,edelman_cover_2022,farah_things_2023,ray_non-parametric_2023}, whereas we only place one peak in the models shown in Figure~\ref{fig:cartoon}.
The peak at $\sim10\Msun$ creates bands in all panels that have very little vertical extent because the peak is proximal to the minimum BH mass. 
Nonetheless, fits using both models exhibit vertical and horizontal bands, indicating peaks in both $p_1(m_1)$ and $p_2(m_2)$.
While the \default{} model requires this behavior, the \diff{} model does not, meaning that the secondary mass distribution appears to exhibit its own feature at $\sim35\Msun$.
Furthermore, the fact that \diff{} produces a PPD similar to that of \default{} indicates that the data support consistent primary and secondary mass distributions. 

The bands in all panels do not go the full extent of parameter space but rather taper off away from the diagonal, indicating a preference for equal-mass binaries either through a pairing function or a mass ratio distribution that favors $m_1 \approx m_2$.

\begin{figure}
    \centering
    \includegraphics[width=\columnwidth]{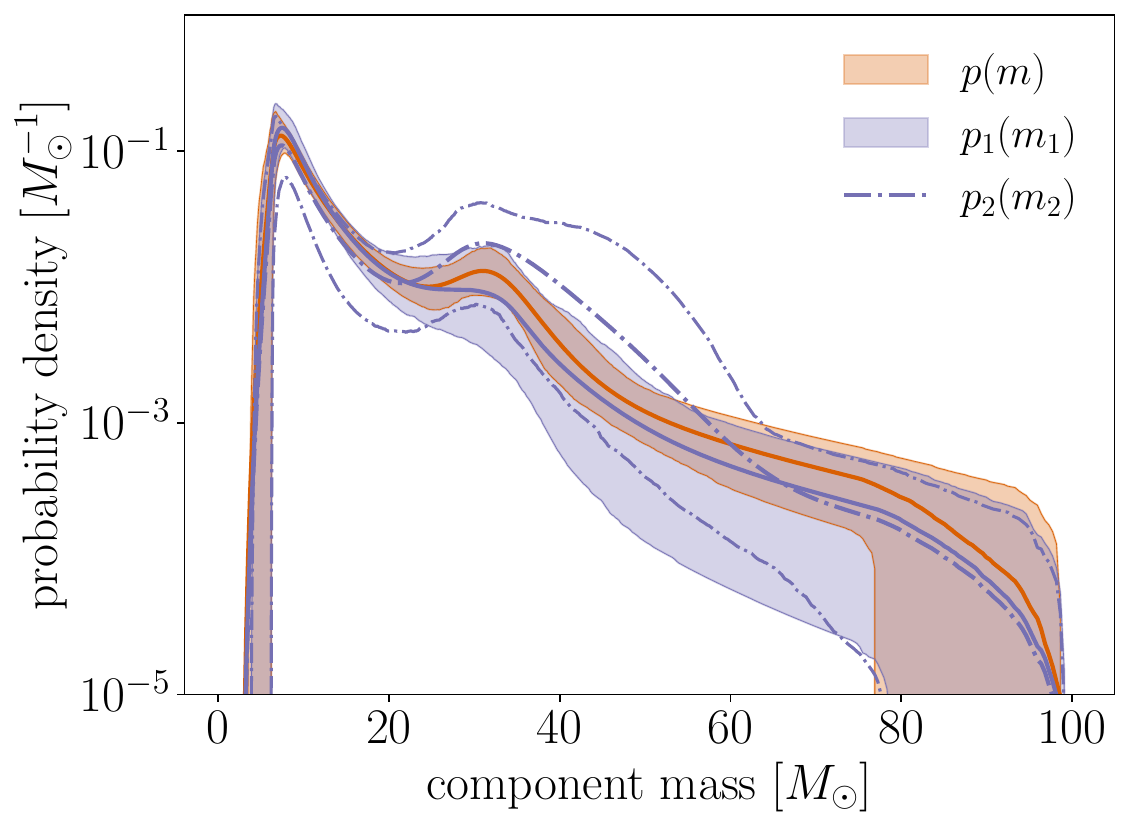}
     \caption{Underlying distribution of primary and secondary masses for the \default{} (\emph{orange}) and \diff{} (\emph{purple}) models.
    Under the \default{} model, $p_1(m_1)=p_2(m_2) \equiv p(m)$, so only $p(m)$ is shown.
    $p(m)$ under \default{} is more tightly constrained than $p_1(m_1)$ (\emph{purple filled band}) or $p_2(m_2)$ (\emph{dot-dashed lines}) under \diff{} as it has twice as many observations per free parameter.
    These distributions are constructed to describe the population of black holes \emph{before} the function that pairs components into merging binaries is applied.
    All three underlying distributions are consistent with one another, though $p_2(m_2)$ appears to have a large peak at $\sim35\Msun$ while $p_1(m_1)$ has some support for no peak in that region.
    $p(m)$ does not have support for no peak, but its peak is constrained to be small, while the peak in  $p_2(m_2)$ is less well-constrained and may be larger in amplitude.
    This hints at the possibility that the peak identified in the primary mass distribution by the \qdist{} formalism may have been driven by a peak in $p_2(m_2)$ rather than $p_1(m_1)$, though hyper-parameter uncertainties within \diff{} are too large to definitively determine this, and the data is consistent with $p_1(m_1) = p_2(m_2)$.
    }
    \label{fig:bucket distributions}
\end{figure}

\subsection{Primary and secondary masses are consistent with having the same underlying distribution}
\label{sec:diff vs same}
The underlying distributions (i.e. before a pairing function is applied) of the primary and secondary masses are shown in Figure~\ref{fig:bucket distributions} for the two pairing function models.
There is a region of overlap between the primary and secondary mass distributions under the \diff{} model, indicating that the primary and secondary mass distributions are consistent with one another.
As expected, this region of overlap also coincides with $p(m)$, the distribution describing both component masses in the \default{} model.

\begin{figure}
    \centering
    \includegraphics[width=\columnwidth]{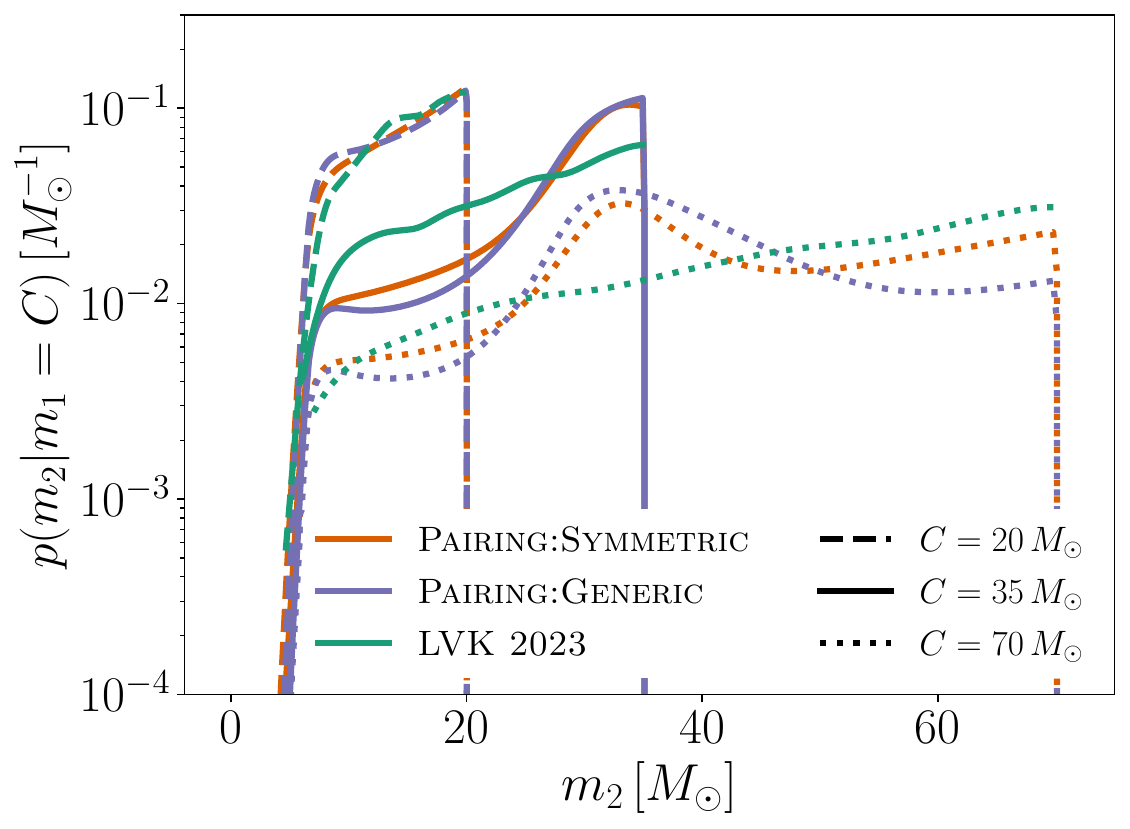}
    \caption{Conditional PPDs for the models considered in this work.
    We show the secondary mass distributions conditioned on $m_1=20\Msun$ (\emph{dashed}), $m_1=35\Msun$ (\emph{solid}) and $m_1=70\Msun$ (\emph{dotted}) to exemplify when the primary mass is below, inside, and above the Gaussian peak, respectively.
    Orange lines correspond to \default, purple lines to \diff, and green lines to \qdist.
    Lines denote the mean posterior probability, marginalized over hyper-parameter uncertainty, and credible intervals are omitted for clarity.
    When $m_1$ is below the peak, all models agree.
    When $m_1$ is above or within the peak, \default{} and \diff{} exhibit a larger preference for $m_2$ to be in the peak than the \qdist{} model does because the latter is constructed to behave like a power law in the range $[m_{\min}, m_1]$.
    The fact that \diff{} and \default{} approximately agree on the peak location and height indicates that the primary and secondary masses may follow the same underlying distribution up to a pairing function.
    }
    \label{fig:conditional ppds}
\end{figure}

Models not explicitly parameterized in terms of a pairing function are unable to produce underlying distributions such as those shown in Figure~\ref{fig:bucket distributions}, so for the sake of comparison to previous work, we turn to conditional\footnote{While marginal distributions are typically used for the purpose of comparing PPDs from several models on a single plot, they are not sensitive to differences between two-dimensional mass distributions when equal masses are preferred (see Appendix~\ref{ap:model comparison}), so we use conditional distributions instead.}
$m_2$ distributions, $p(m_2|m_1=C)$, where $C$ can be any number in the domain of the $m_1$ distribution.
Figure~\ref{fig:conditional ppds} shows these conditional distributions for the \default{}, \diff{} and \qdist{} models, the latter of which does not use a pairing function.
The curves in Figure~\ref{fig:conditional ppds} are averaged over the hyper-posterior for each model. 

For all values of $C$, the inferred distribution under the \diff{} model behaves similarly to that of the \default{} model, indicating consistency between the primary and secondary mass distributions.
In particular, under the \default{} and \diff{} models, we see a peak in the conditional secondary mass distribution $p(m_2 | m_1 = C)$ when $C \geq 35\,M_\odot$ (solid and dotted lines). 
While $m_2$ generally prefers to be near $m_1$, as indicated by an upward trend in all models, the two pairing function models have more support for $m_2$ being within the peak than being nearly equal to $m_1$ when $m_1$ is large (dotted lines).
This behavior is in contrast to the \qdist{} model, which forces the conditional secondary mass distribution $p(m_2 | m_1 = C)$ to monotonically increase for all values of $C$ because it does not allow for a peak in $m_2$.
While the \diff{} model can replicate this behavior, it instead recovers similar distributions to the \default{} model.
The Bayes factor of \diff{} relative to \default{} is \result{$\mathcal{B}($\diff$)/\mathcal{B}(\default) = 0.38$}, indicating an inability to distinguish between the two models, with perhaps a modest preference for the \default{} model.

\subsection{Improved constraints on peak location}

\begin{figure}
    \centering
    \includegraphics[width=\columnwidth]{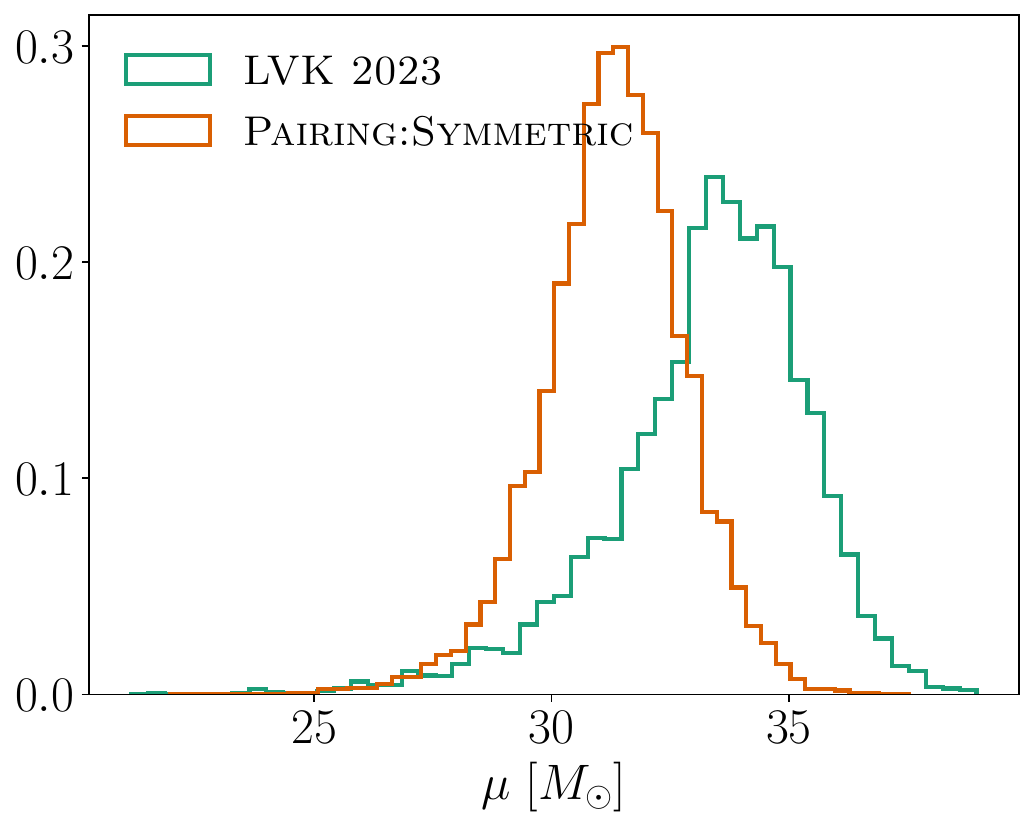}
    \caption{Marginal posteriors on $\mu$, the location of the Gaussian peak, for the \qdist{} and \default{} models.
    This parameter is more precisely measured under the \default{} formalism by \result{$\sim1/\sqrt{2}$}, indicating that $m_1$ and $m_2$ independently contribute to this measurement.
    $\mu$ is the most correlated parameter with the Hubble constant when using ``spectral sirens'' \citep{ezquiaga_spectral_2022,abbott_constraints_2023}, so using the \default{} model will presumably improve cosmological measurements. 
    The central values of the two distributions differ because the hyper-parameter has slightly different effects on the resulting 2D mass distribution each model.}
    \label{fig:mpp marginal}
\end{figure}

If the primary and secondary mass distributions are identical, we can better measure the properties of features in that common distribution by using measurements of both $m_1$ and $m_2$.
Figure~\ref{fig:mpp marginal} shows the one-dimensional hyper-posterior for the location, $\mu$, of the Gaussian peak in the mass distribution under the \default{} and \qdist{} models.
We measure $\mu$ with a standard deviation of \result{$1.50\Msun$} under the \default{} model compared to \result{$2.08\Msun$} under the \qdist{} model, representing a \result{28\%} improvement.
This increase in precision is similar to that expected from using twice the number of independent events ($1-1/\sqrt{2}=0.29$), because now both $m_1$ and $m_2$ contribute to the inference, rather than just $m_1$. 

Furthermore, the \default{} model recovers a lower value of \result{$\mu = 31.4_{-2.6}^{+2.3}\Msun$} (median and 90\% credible interval) compared to the \qdist{} result of \result{$\mu = 33.6_{-4.0}^{+2.6}$}, because it refers to a feature in the underlying $p(m)$ distribution rather than the marginal $m_1$ distribution. Features in $p(m)$ appear at larger masses in the marginal $m_1$ distribution and lower masses in the marginal $m_2$ distribution due to the constraint that $m_1 > m_2$.

If the $\sim35\Msun$ peak is believed to be a feature of the supernova remnant mass distribution, it is best to use its location in the underlying $p(m)$ distribution rather than its location in the marginal distribution.
For example, analyses wishing to compare this feature with expectations from a pair-instability supernova pileup~\citep[e.g.][]{stevenson_impact_2019,farmer_constraints_2020} should use the value presented here (\result{$\mu = 31.4_{-2.6}^{+2.3}\Msun$}). 
Interestingly, this lower peak location is in further tension with the latest theoretical predictions for a pair-instability pileup \citep{farag_resolving_2022}.  

When possible, though, it is best to compare theoretical predictions directly to the two-dimensional mass distribution (such as those in Figure~\ref{fig:2d ppds}) rather than to the values of specific hyper-parameters, since hyper-parameters have different meanings in different models.
Correspondingly, it is important that models used to fit the data are intentionally constructed to allow for features in both the primary and secondary mass distribution. 
Many of the current parametric \citep{fishbach_where_2017,talbot_measuring_2018,baxter_find_2021,abbott_population_2021,abbott_population_2023} and non-parametric \citep{tiwari_vamana_2021,edelman_aint_2022,callister_parameter-free_2023,godfrey_cosmic_2023} BBH mass distribution models enforce asymmetry between $m_1$ and $m_2$, excluding the possibility that the primary and secondary mass distributions share the same features\footnote{\citet{callister_parameter-free_2023} and \citet{godfrey_cosmic_2023} use flexible models to fit $p(q)$ rather than using power laws, but they do not use a pairing function and instead fit the $m_1$ and $q$ distributions independently, therefore assuming the product of the marginal distributions equals the two-dimensional distribution.
This is the same fundamental choice as is described in Equation~\ref{eq:mass ratio dist}, and therefore assumes that $m_1$ and $m_2$ follow different distributions.} (with a few exceptions, e.g. \citealt{edelman_cover_2022, ray_non-parametric_2023,sadiq_binary_2023}).
\new{It is possible to construct a mass distribution that allows for features in both the primary and secondary mass distribution without using a pairing function \citep[e.g.][]{ray_non-parametric_2023}, but directly comparing the primary and secondary mass distributions using these parametrizations is less straightforward.}

Another application of our improved measurement of the peak location is ``spectral siren'' cosmology, which uses such features in the mass distribution to infer the expansion history of the universe \citep{chernoff_gravitational_1993,messenger_measuring_2012, taylor_cosmology_2012,farr_future_2019, ezquiaga_jumping_2021,ezquiaga_spectral_2022,abbott_constraints_2023}.
Previous analyses found the location of this peak to be the parameter most correlated with the Hubble constant \citep[see Figures 5 and 13 of][]{abbott_constraints_2023}, so an improved constraint on $\mu$ should therefore have a relatively large effect on the constraints of cosmological parameters if all other mass distribution parameters remain equally-well constrained.



\subsection{Evidence for a peak in the secondary mass distribution}
\label{sec:peak in m2}

Consistency between the primary and secondary mass distributions implies a peak in the secondary mass distribution at $m_2\sim30\Msun$, as this feature has already been robustly identified in the primary mass distribution~\citep{abbott_population_2021,tiwari_vamana_2021,abbott_population_2023,edelman_cover_2022,callister_parameter-free_2023,farah_things_2023}.
However, the data is also consistent with differing primary and secondary mass distributions, as shown by the regions in which the filled and dashed bands do not overlap in Figure~\ref{fig:bucket distributions}.
In this case, it is worthwhile to explicitly determine whether there exists a peak in the secondary mass distribution.

\begin{figure*}
     \centering
     \subfloat{
         \centering
         \includegraphics[width=0.48\textwidth]{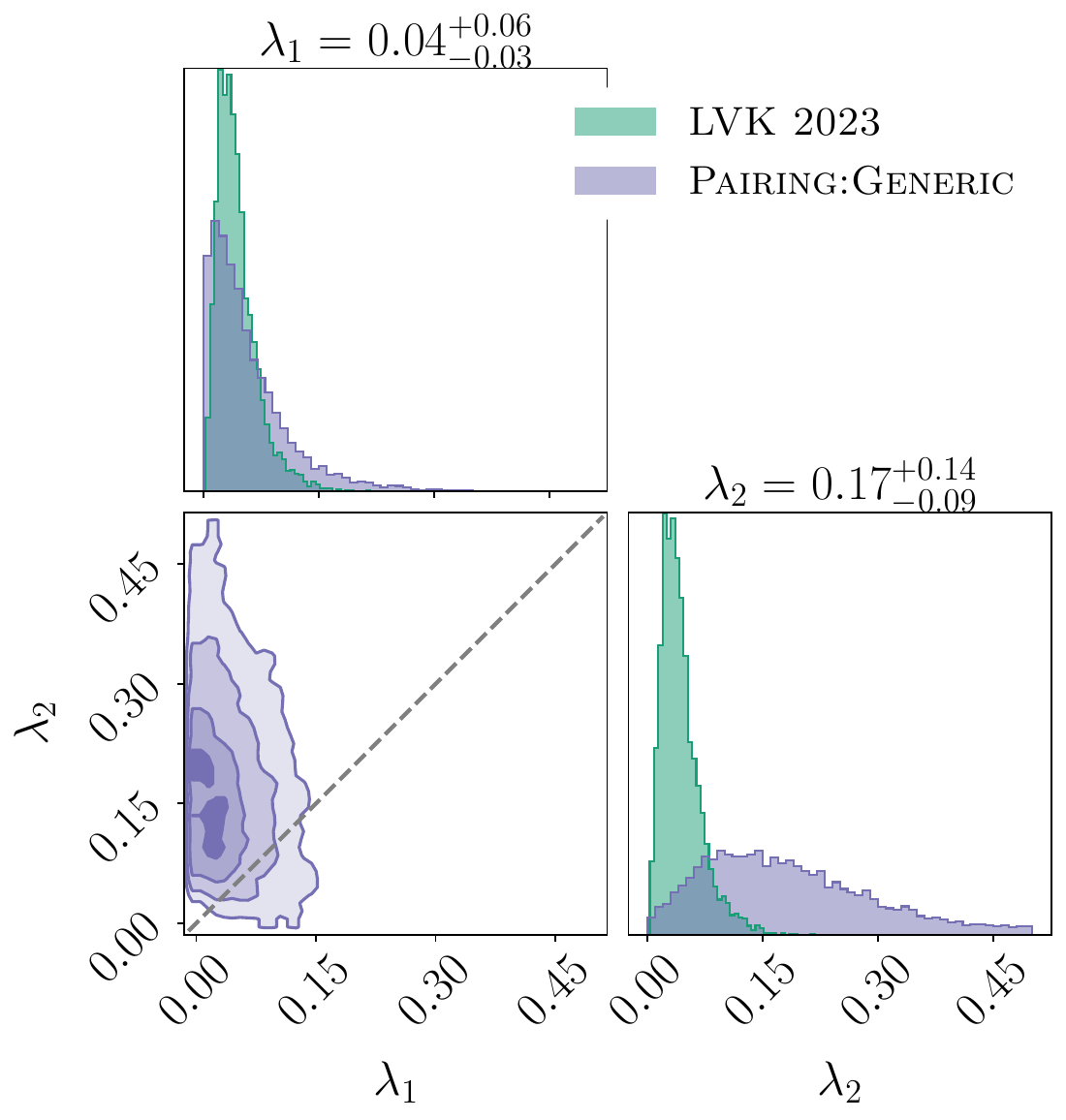}
     }
     \hfill
     \subfloat{
         \centering
         \includegraphics[width=0.48\textwidth]{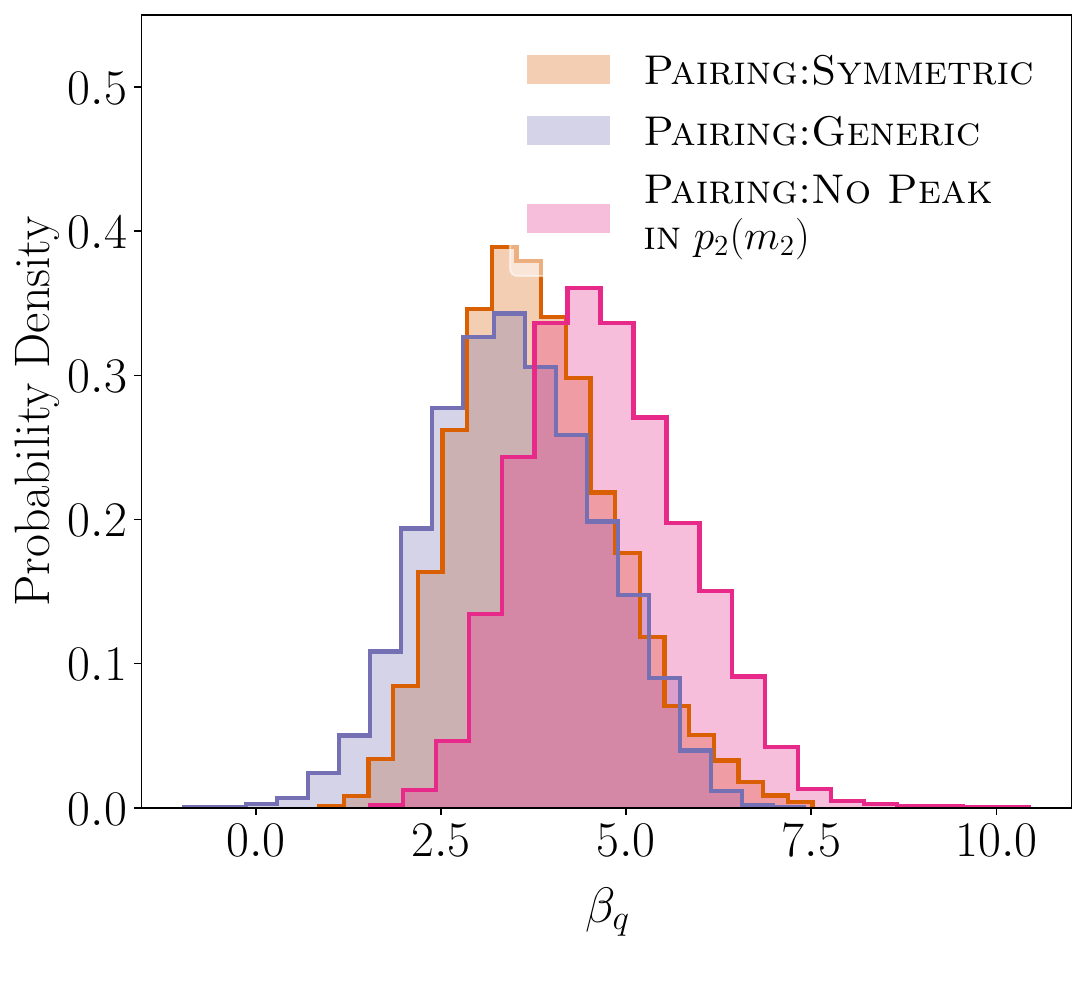}
     }
    \caption{Hyperposteriors under pairing function models.
    \emph{Left:} Corner plot of hyper-parameters governing the height of the Gaussian peak for the primary ($\lambda_1$) and secondary ($\lambda_2$) mass distributions under the \diff{} model.
    The medians and 90\% credible intervals are shown above their respective marginal distributions.
    The medians indicate that roughly 17\% of BBHs have secondary masses in the Gaussian peak while 4\% have primary masses in the peak, though this preference for a larger peak in $p_2(m_2)$ may be due to the relatively poor constraint on $\lambda_2$ rather than a true preference in the data.
    The lower left portion of the two-dimensional posterior is excluded, indicating that there must be a peak in either $p_1(m_1)$ or $p_2(m_2)$, and a slight preference exists for the peak to be in $p_2(m_2)$ (upper left portion of plot) or both distributions (central portion of plot), though the peak being in $p_1(m_1)$ only is not completely ruled out.
    For reference, the 1D posterior on $\lambda$ under the \qdist{} model is shown in green, with $\lambda=0.04^{+0.03}_{-0.02}$ \citep{abbott_population_2023}.
    The dashed grey line indicates where $\lambda_1=\lambda_2$.
    \emph{Right:} Marginal posteriors on $\beta_q$, the hyper-parameter controlling the steepness of the pairing function under the \default, \diff, and \lamTwoeqZero{} models.
    When we set $\lambda_2=0$, $\beta_q$ increases, indicating a preference for equal-mass components.
    This behavior is likely caused to accommodate an excess of events with $m_2\sim35\Msun$, which can either be caused by a peak in $p_2(m_2)$ or with a peak in $p_1(m_1)$ and a strong preference for equal-mass binaries (see discussion in Appendix~\ref{sec:pairing-vs-pq-pedagogy}).
}
    \label{fig:lam1 lam2 corner}
\end{figure*}

The left panel of Figure~\ref{fig:lam1 lam2 corner} shows the posterior distribution for the parameters governing the height of the peak in $m_1$ and $m_2$ under the \diff{} model, $\lambda_1$ and $\lambda_2$, respectively. 
Setting $\lambda_{\{1,2\}}=0$ means that no binaries in the astrophysical population are in the Gaussian peak\footnote{The integrated fraction of events in the region near the peak is higher than the fraction indicated by $\lambda$ because both the Gaussian peak and the underlying power law contribute to the rate in that region. Therefore, $\lambda_{\{1,2\}}=0$ does not mean that there are no events with masses $\sim35\Msun$.}, while $\lambda_{\{1,2\}}=1$ means all binaries are in the Gaussian peak.
The lower left and upper right regions of the two-dimensional posterior are excluded, meaning that $\lambda_1$ and $\lambda_2$ cannot both be zero, nor can they both be large.
This indicates that either one of the two distributions has a large peak while the other has none, or that both distributions have moderate peaks.

In fact, the posterior on $\lambda_2$ peaks at a higher value than $\lambda_1$.
Under the \qdist{} model, the 1D posterior on $\lambda$ peaks between the marginal $\lambda_1$ and $\lambda_2$ posteriors of the \diff{} model.
Additionally, in the \diff{} model, $\lambda_1$ has more support at zero relative to $\lambda$, while $\lambda_2$ has reduced support at zero. 
This hints at the intriguing possibility that the secondary masses may be driving the nonzero value of $\lambda$ found by \citet{abbott_population_2021} and \citet{abbott_population_2023}.
In other words, \emph{it is possible that the $\sim30\Msun$ peak exists the secondary mass distribution rather than the primary mass distribution.} 
However, the data are still consistent with $\lambda_1=\lambda_2$: the dashed grey line in Figure~\ref{fig:lam1 lam2 corner} intersects the $1.5$-$\sigma$ contour of the hyper-posterior.

As shown in Appendix~\ref{ap:all_same_lam2_0}, the \qdist{} formalism behaves similarly to \lamTwoeqZero{}, which is nested within \diff{} when $\lambda_2=0$.
Therefore, ruling out $\lambda_2=0$ would indicate that the pairing function formalism is strongly preferred.
We find that $\lambda_2=0$ is disfavored but not ruled out: there is \result{4.2} times more posterior density at $\lambda_2=0.17$ (its median \emph{a posteriori} value) than at $\lambda_2=0$.
It is therefore difficult to tell with the current number of detections whether the data prefer for the peak to exist in $p_1(m_1)$, $p_2(m_2)$, or both.

These three possibilities are somewhat degenerate because of the preference for equal mass BBHs.
To illustrate this, the right panel of Figure~\ref{fig:lam1 lam2 corner} compares the power law spectral index, $\beta_q$, of the pairing function under \diff{} and \lamTwoeqZero.
When there is no peak in $p_2(m_2)$, the $\beta_q$ posterior shifts to higher values, which corresponds to a larger preference for $q\sim1$.
This is because the \lamTwoeqZero{} model has a peak in $p_1(m_1)$, so it allows for a high fraction of secondary masses to lie within the peak by making more binaries equal-mass.
For comparison, we also plot the posterior on $\beta_q$ under \default{} to show how a different set of assumptions about $\lambda_2$ changes $\beta_q$.
The shift in $\beta_q$ when $\lambda_2=0$ is larger than when $\lambda_2=\lambda_1$, suggesting that it is driven by an excess of events with $m_2\sim35\Msun$ rather than by other model assumptions or a different realization of the inference.

In summary, we find modest evidence for a peak in $p_2(m_2)$, suggesting that pairing function models may be preferred over the \qdist{} formalism, though current observations do not allow us to definitively rule out that the peak exists only in $p_1(m_1)$. 
The data is consistent with $\Lambda_1=\Lambda_2$, so there is not strong evidence against the possibility that $m_1$ and $m_2$ are drawn from the same underlying distribution, up to a pairing function.

\subsection{Binary black holes are picky}
\label{sec:pickiness}
Pairing functions provide an intuitive way to quantify whether the properties of one black hole in a binary influences those of its companion. 
Binaries that pair up independently of each component's masses are described by a pairing function with $\beta_q = 0$, whereas a preference for equal-mass binaries is described by $\beta_q > 0$.
We find $\beta_q>0$ to \result{$>99.99$}\% for \default{} and to \result{99.95}\% for \diff. 
This is consistent with an earlier study by \citet{fishbach_picky_2020} who use the pairing function formalism on GWTC-1 \citep{abbott_gwtc-1_2019}.
However, BBHs are not maximally picky: the posterior on $\beta_q$ is not railing against the high end of the prior bounds.

Our inferred two-dimensional mass distribution is generally consistent with that shown in \citet{sadiq_binary_2023}, who use a non-parametric approach and assume that the population is symmetric under $m_1\leftrightarrow m_2$.
They show evidence for peaks in the mass ratio distribution at $q\sim0.5$ when $m_1\sim15\Msun$ and $m_1\sim70\Msun$, which they interpret as a lack of preference for similar-mass pairings.
However, we note that these peaks in the mass ratio distribution translate to peaks in the secondary mass distribution at $\sim 7\Msun$ and $\sim35\Msun$, the same locations at which the primary mass distribution exhibits peaks.
Parametrizations that assume symmetry under $m_1\leftrightarrow m_2$ but do not use a pairing function are unable to disentangle the effects of preference for similar-mass pairings from features in one or both component mass distributions.
We therefore conclude that both our results and those presented in \citet{sadiq_binary_2023} are consistent with a preference for similar-mass pairings as well as structure in the secondary mass distribution.

In this work, we have only examined power law forms for the pairing function, but more complex models are under investigation and will be presented in future work.
If the pairing function is a more complex function of mass ratio, or if it depends on multiple parameters, such as primary mass or spin \citep[e.g.][]{farah_bridging_2022}, it may become easier to distinguish between pairing function models and models parameterized similarly to \qdist{}, as complex pairing functions may find more support away from the $m_1=m_2$ diagonal.
The same is true if the marginal mass ratio distribution has multiple modes, or if mass ratio is allowed to correlate with other parameters \citep[e.g. effective spin,][]{callister_who_2021}.

\section{Looking to the future}
\label{sec:future}
The \qdist{} and pairing function models differ most in their predictions for the number of events in the region $m_1\in[35,40] \cap m_2\in[m_{\min},30]$, because they disagree on whether to place $m_2$ in the Gaussian peak or somewhat evenly throughout the available parameter space.
We can therefore determine which model will be preferred in the future by counting the number of detected events in that region.

It is expected that $260^{+330}_{-150}$ BBHs will be detected by the end of the LVK's fourth observing run (O4), and $870^{+1100}_{-480}$ BBHs will be detected by the end of the fifth observing run (O5) \citep{Kiendrebeogo_updated_2023}.
With 260 (870) total BBH detections, we expect \result{$23.6\pm 4.6$ ($79.0 \pm 8.5$)} BBHs to fall in the region $m_1\in[35,40] \cap m_2\in[m_{\min},30]$ under the \qdist{} model and \result{$12.6\pm 3.5$ ($42.3 \pm 6.3$)} under the \default{} model.
This means that by the end of O4, we will be able to distinguish the \qdist{} and \default{} models to $>2\sigma$, and we will be able to distinguish them to $>4\sigma$ by the end of O5.

Of course, it will be necessary to perform a full hierarchical analysis, as measurement uncertainties of detected systems will cause events to scatter into and out of this region. 
Additionally, in a hierarchical Bayesian context, posterior predictive checks performed on the event-level parameters (such as the true masses of individual events) are less sensitive than those performed on the observed data, \new{such as the strain or the maximum likelihood event parameters} \citep{sinharay_posterior_2003,gelman_prior_2006,bayarri_bayesian_2007,loredo_bayesian_2013}.
This is why, for example, \citet{fishbach_most_2020} performs posterior predictive checks on the maximum-likelihood values of the observed and predicted events \new{rather than the posterior distributions of those events}.
Therefore, more sensitive posterior predictive checks may be able to distinguish between the two frameworks with fewer events than we project here.

\section{Summary and Implications for Astrophysics of Merging BBHs}\label{sec:discussion}
\new{
We present the first analysis of the secondary mass distribution of merging BBHs.
This allows us to explore whether the primary and secondary component masses in merging BBHs are drawn from the same distribution, or whether the BBH mass distribution is asymmetric in $m_1 \leftrightarrow m_2$ as is commonly assumed in BBH population studies.
We find the data to be consistent with two possibilities: either the primary and secondary mass distributions are similar, or a larger peak exists in the secondary mass distribution than the primary mass distribution.
In either scenario, a peak likely exists in the secondary mass distribution.
This possibility is not considered in many previous analyses of the BBH population, which fix the secondary mass distribution to be a power law \citep[e.g.][]{abbott_binary_2019, abbott_population_2021, abbott_population_2023} or assume that $m_1$ and $m_2$ are interchangeable \citep[e.g.][]{fishbach_picky_2020, sadiq_binary_2023}.
The existence of this secondary-mass peak has implications for the formation channels of merging BBHs and the origin of the $\sim35\,\Msun$ peak.
}

If the mass distribution is indeed symmetric under $m_1 \leftrightarrow m_2$ as in our preferred model \default{}, this may imply that a large fraction of merging BBHs are formed through dynamical assembly, in which the two component BHs are born through a similar process and then find each other in a dense stellar environment.
In this case, the pairing function and its dependence on mass ratio and/or total mass may encode valuable information about dynamical processes like mass segregation and binary exchanges. 

The appearance of a peak in both primary and secondary mass distributions is consistent with an origin in the BH remnant mass function. 
However, the peak location at \result{$31.4^{+2.3}_{-2.6}\,\Msun$} is in tension with predictions for pair-instability supernovae, \new{so this feature is likely caused by another astrophysical process}.
As another application of our work, we suggest that future spectral siren measurements consider the \default{} model when inferring the Hubble constant, because it provides an improved measurement of the peak location relative to the \qdist{} model and the peak location is strongly correlated with the inferred value of $H_0$.

On the other hand, the data remain consistent with an asymmetric BBH mass distribution \p{}, in which $p_1(m_1)\neq p_2(m_2)$, as in the \diff{} model. 
This would imply that a significant fraction of merging BBHs originate from field binaries, in which ``primary" and ``secondary" are physically meaningful labels if they tend to correspond to the first-born or second-born BH in a binary.
The component mass distributions $p_1(m_1)$ and $p_2(m_2)$ would then encode the binary IMF and the highly uncertain physics of binary stellar evolution.
Specifically, the location and prominence of features in each component mass distribution may provide insight into how mass loss versus mass accretion affect stellar evolution and collapse, including their effect on supernovae and the BH remnant mass function.
\new{For example, when mass transfer is unstable, a peak is expected at $\sim15\Msun$ in the primary mass distribution, but near the minimum black hole mass in the secondary mass distribution, though the relative locations of these peaks depends on common-envelope and supernova kick physics \citep{van_son_redshift_2022}.}

Notably, if the BBH mass distribution is asymmetric, we find that \emph{it is possible that no peak exists in the primary mass distribution, and the previously-identified peak in primary mass is actually driven by an overdensity in the secondary mass distribution}.
A peak that is more prominent in $p_2(m_2)$ than $p_1(m_1)$ may imply that pulsational pair-instability supernovae are more frequent or are shifted to lower masses among second-born BHs.

The degree of asymmetry in the BBH mass distribution provides insight into the frequency of mass inversion in binary stars, providing a complementary probe to BH spins~\citep{mould_which_2022}.
If mass inversion never occurs, $m_1$ will typically be the remnant of the donor star.
However, if mass inversion is the norm, $m_1$ will be the remnant of the accretor star.
A perfectly symmetric BBH distribution under isolated binary evolution would imply that mass inversion happens exactly 50\% of the time, causing the primary and secondary mass distributions to be indistinguishable even though binary physics imparts different distributions on the first-born versus second-born black hole, but this is statistically unlikely.
\new{Determining whether primary and secondary BH masses follow the same underlying distributions is therefore a novel and promising probe of formation channels.
However, this test should be interpreted within the full BBH population inference, as theoretical models should make consistent predictions for the mass, spin and redshift distributions.}

\new{We have only explored parametric models in this work, and can therefore only comment on the features in the mass distribution explicitly parameterized by our model.
Future investigations may find it beneficial to use a non-parametric approach that separately fits the primary and secondary mass distributions by employing a pairing function.}



We additionally find that BHs in merging binaries have a strong preference to pair with similar-mass BHs for all forms of the secondary mass distribution considered in this work.
This is consistent with the results presented in \citet{fishbach_picky_2020}, though we more confidently exclude the scenario in which BBHs pair independently of mass ratio since we now have many more detected BBHs.

By the end of the LVK's \result{fifth} observing run, we expect to confidently distinguish between the different scenarios presented here for the BBH mass distribution. In this work, we found that secondary masses in merging BBH systems likely display a peak at $\sim35\,M_\odot$, whereas previous results identified this peak exclusively among primary masses. 
With a few hundred additional BBH observations, we expect to determine whether both component mass distributions have a peak at the same location and if the peak is more prominent among secondary masses. 


\new{Directly incorporating the distribution of secondary masses can serve as an important tool to constrain formation mechanisms of BBHs.}

\begin{acknowledgments}
The authors thank Sharan Banagiri, Thomas Dent, Zoheyr Doctor, Will Farr, Davide Gerosa, Tom Loredo, and Jam Sadiq for helpful discussions. 
A.M.F. is supported by the National Science Foundation Graduate Research Fellowship Program under Grant No. DGE-1746045.
D.E.H is supported by NSF grants AST-2006645 and PHY-2110507, as well as by the Kavli Institute for Cosmological Physics through an endowment from the Kavli Foundation and its founder Fred Kavli.
This research has made use of data or software obtained from the Gravitational Wave Open Science Center (gwosc.org), a service of the LIGO Scientific Collaboration, the Virgo Collaboration, and KAGRA.
This material is based upon work supported by NSF's LIGO Laboratory which is a major facility fully funded by the National Science Foundation.
The authors are grateful for computational resources provided by the LIGO Laboratory and supported by National Science Foundation Grants PHY-0757058 and PHY-0823459.
\end{acknowledgments}

\vspace{5mm}
\facilities{LIGO, Virgo}

\software{\texttt{gwpopulation} \citep{talbot_parallelized_2019}, \texttt{bilby} \citep{ashton_bilby_2019}, \texttt{scipy} \citep{virtanen_scipy_2020}, \texttt{corner.py} \citep{foreman-mackey_cornerpy_2016}
          }

\appendix
\renewcommand{\qdist}{\textsc{Conditioned-q}}
\section{Comparison of Common Mass Distribution Parametrizations}
\label{ap:conditioned q}

\subsection{Differences between mass ratio distributions and pairing functions}
\label{sec:pairing-vs-pq-pedagogy}
Given a form for the primary mass distribution, there are several ways to construct a two-dimensional mass distribution.
We discuss two possibilities here that are common in the literature, showing the different effects they have on the resulting two-dimensional mass distribution.

\begin{figure*}
    \centering
    \includegraphics[width=\textwidth]{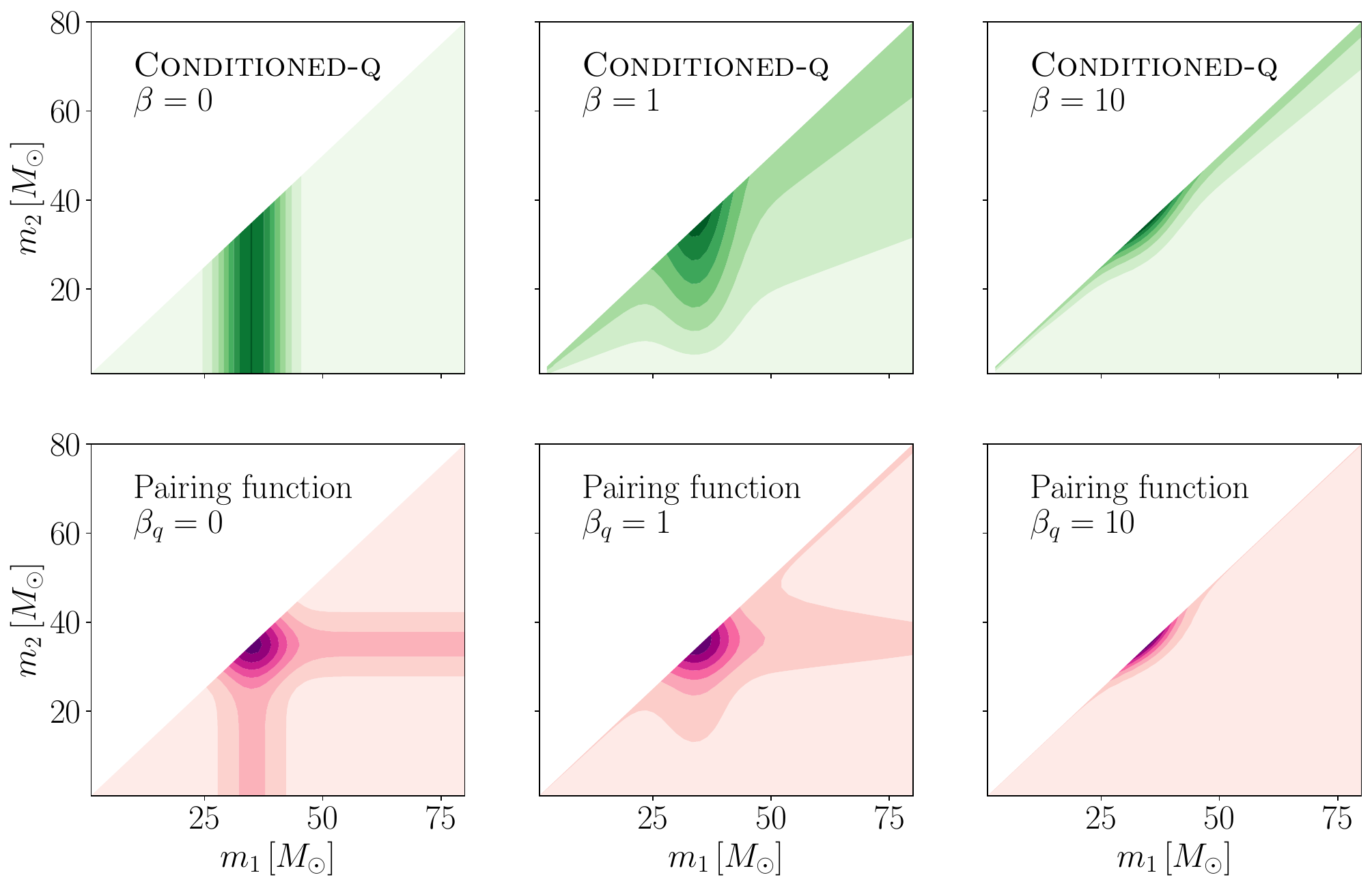}
    \caption{Illustration of some possible two-dimensional mass distributions under the commonly used ``\qdist{}'' formalism described by Equation~\ref{eq:mass ratio dist} (\emph{top row}) and the pairing function formalism described by Equation~\ref{eq:pairing function} (\emph{bottom row}).
    Overdensities/peaks in the mass distribution appear as darker filled contours in these figures.
    The \qdist{} formalism is only able to produce models with features in the $m_1$ distribution, as shown by the vertical bands in the top row, whereas the pairing function formalism can model features in either $m_1$ or $m_2$, or both.
    The different columns correspond to different power law spectral indices for the mass ratio distribution (\emph{top row}, $\beta$) and mass ratio-dependent pairing function (\emph{bottom row}, $\beta_q$).
    In the case of a steeply rising mass ratio distribution, or if components strongly prefer to pair with nearly equal-mass partners, the \qdist{} model and the pairing function model become difficult to tell apart and likely explain the data equally well, as shown by the two panels in the leftmost column.
    The diagonal contours in the middle and right columns are caused by a preference for equal-mass binaries and follow lines of constant mass ratio.
    }
    \label{fig:cartoon2}
\end{figure*}

The main qualitative differences between the two parametrizations are illustrated in Figure~\ref{fig:cartoon2}.
The top row has three examples of mass distributions that can be described by a model of the form
\begin{equation}
    p(m_1,m_2|\Lambda_m,\beta) = p(m_1|\Lambda_m)p(q|m_1,m_{\min},\beta)
    \label{eq:mass ratio dist}
\end{equation}
for different values of $\beta$. 
Here, $m_1$ is the mass of the heavier component in the binary, $m_2$ is the mass of the lighter component, and $q=m_2/m_1 \leq 1$ is the mass ratio.
This can be equivalently written as
\begin{equation}
    p(m_1,m_2|\Lambda_m,\beta) = p(m_1|\Lambda_m)p(m_2|m_1,\beta),
\end{equation}
since $m_2 = q m_1$.
The parametrization described in Equation~\ref{eq:mass ratio dist} is used by all parametric models presented in \citet{abbott_population_2021} and \citet{abbott_population_2023} that were used to model the primary mass distribution of BBHs, such as \textsc{Broken Power Law} and \textsc{Power Law + Peak}, and is also commonly used in other analyses \citep[e.g.,][]{fishbach_where_2017,kovetz_black_2017,talbot_measuring_2018,tiwari_vamana_2021,edelman_aint_2022,callister_parameter-free_2023,godfrey_cosmic_2023}.
For the remainder of this Appendix, we will refer to models parameterized by Equation~\ref{eq:mass ratio dist} as ``\qdist,'' since they require the mass ratio distribution to be explicitly conditioned on primary mass.

The bottom row of Figure~\ref{fig:cartoon2} has three examples of mass distributions that can be described using a ``pairing function'', $f$ \citep{fishbach_picky_2020}.
Models with pairing functions have the form
\begin{equation}
    p(m_1,m_2|\Lambda) = p_1(m_1|\Lambda_1)p_2(m_2|\Lambda_2)f(q;\beta_q)
    \label{eq:pairing function}
\end{equation} 
where $f(q)$ is a pairing function that depends on the mass ratio of the system\footnote{In principle, the pairing function can be parameterized in terms of any observable parameter (e.g., total mass).
}, and $\Lambda = \{\Lambda_1, \Lambda_2, \beta_q\}$ is the set of all model hyper-parameters.
In this work, we use a pairing function of the form $f(q;\beta_q) = q^{\beta_q} \Theta(q\leq1)$, though other forms may provide a better fit to the data \citep[e.g.][]{farah_bridging_2022}. 
In the examples illustrated in Figure~\ref{fig:cartoon2}, the primary and secondary mass distributions are equivalent, so $p_1(m) = p_2(m) \equiv p(m)$.
Alternatively, $\Lambda_1 = \Lambda_2$.
This describes a situation in which there is a single underlying mass distribution from which both components are drawn.
The pairing function then describes how likely the two components are to be combined in a merging binary based on their mass ratio.
A pairing function that prefers equal mass binaries will cause a marginal mass ratio distribution that has more support near $q=1$, but the inverse is not necessarily true.

The parametrization in Equation~\ref{eq:pairing function} factorizes the possibility that components in binaries prefer to be near-equal mass and the possibility that the primary and secondary masses have distinct probability distributions.
In other words, pairing functions allow us to model the secondary mass separately from the primary mass, while also allowing for the possibility that component BHs prefer to pair with similar-mass BHs.

In Figure~\ref{fig:cartoon2}, a peak at $35\Msun$ is placed in both models to show the effects of such features in both cases.
One consequence of \qdist{} models is that features such as Gaussian peaks can only appear in the the primary mass distribution.
This is shown by the vertical bands in the top row of Figure~\ref{fig:cartoon2} and lack of horizontal bands, since a band in the vertical (horizontal) direction are caused by a peak or dip in the primary (secondary) mass distribution for a range of secondary (primary) masses.
For pairing function models, features can appear in $p_1(m_1)$, $p_2(m_2)$, or both.
We have illustrated the case in which the same feature appears in both component mass distributions, and this appears as both the vertical and horizontal bands in the bottom row of Figure~\ref{fig:cartoon2}.
If $\Lambda_1$ and $\Lambda_2$ are allowed to differ, features could appear in only one of these distributions.
This would cause there to only be horizontal bands if features only existed in $p_2(m_2)$, and only vertical bands if features only existed in $p_1(m_1)$.
Features are also able to appear in different locations in $p_1(m_1)$ vs $p_2(m_2)$ under the pairing function formalism.
However, the \qdist{} formalism only allows for bands in the vertical direction, meaning that it is not flexible enough to capture true underlying distributions with features in $p_2(m_2)$.
The behavior of the \qdist{} formalism can in general be approximated by the pairing function formalism, while the opposite is not true.

The different columns in Figure~\ref{fig:cartoon2} correspond to different power law spectral indices for the mass ratio distribution (\emph{top row}, $\beta$) and mass ratio-dependent pairing function (\emph{bottom row}, $\beta_q$).
The top row's leftmost panel has a uniform mass ratio distribution, the middle panel has a mass ratio distribution that mildly favors equal-mass binaries, and the right panel's mass ratio distribution strongly favors equal mass binaries.
Analogously, the bottom row's leftmost panel shows a model where components in the binary are allowed to pair up independently of mass ratio, the middle panel shows a model where components have a slight preference to pair up with partners that are equal mass, and the rightmost panel shows the case where components are ``picky'': they almost always pair up with equal mass partners \citep{fishbach_picky_2020}.
When the mass ratio distribution is broad, or when components pair near-independently of mass ratio, the \qdist{} models produce noticeably different distributions than the pairing function models.
However, in the case of very picky binaries or a steeply rising mass ratio distribution, the \qdist{} and pairing function models become difficult to tell apart, and likely explain the data equally well.
There is therefore a degeneracy between the steepness of the pairing function and the existence of distinct features in the two mass distributions \citep[see][for a discussion of this phenomenon in terms of Jacobian transformations]{tiwari_whats_2023}.

Fortunately, as we will show in Section~\ref{sec:pickiness}, we measure $\beta\sim3.5$ and $\beta_q\sim1$, so the data lie somewhere between the middle and rightmost columns.\footnote{The value for $\beta$ under \qdist{} is different than the value found for $\beta_q$ under the pairing function models because the two parameters cause different behaviors in the 2D mass distribution within their respective models.
A low value for $\beta$ does not imply that BBHs are not picky, just that the marginal mass ratio distribution rises slowly.
We discuss this in more detail in Appendix~\ref{ap:betas}.}
This means that differentiating between the two scenarios will be difficult, but possible given enough data.

Mass distributions of compact objects are often visualised through a plot of the marginal component mass distributions.
The marginal $m_2$ distribution is defined as 
\begin{equation}
    p(m_2|\Lambda) = \int \mathrm{d}m_1 p(m_1,m_2|\Lambda) ,
    \label{eq:marginal m2}
\end{equation}
where $p(m_1,m_2|\Lambda)$ can be parameterized in the way described in Equation~\ref{eq:mass ratio dist} or Equation~\ref{eq:pairing function}. 
However, in all but the top left panel in Figure~\ref{fig:cartoon2}, the \emph{marginal} secondary mass distribution exhibits a peak between $30\Msun$--$40\Msun$, even though the secondary mass distribution is not able to have any features on its own. 
This is because features in the primary mass distribution induce features in the marginal secondary mass distribution if equal-mass binaries are preferred:
when a binary's primary mass is within the peak, its secondary mass is also likely to be in that peak simply because $m_2\simeq m_1$ is preferred. 
It is therefore difficult to distinguish between these different scenarios by looking at the marginal distributions alone.
Instead, we analyze two-dimensional distributions such as the ones illustrated in Figure~\ref{fig:cartoon2}, as well as the secondary mass distributions conditioned at various values of $m_1$. 
The latter can be thought of as one-dimensional slices of the former.

It is our goal to determine whether the data prefer models described by pairing functions or by \qdist{} functions.
We also aim to determine if the primary and secondary masses are drawn from the same distribution, and whether we can draw physical insights from features that appear in the primary mass distribution, secondary mass distribution, or both.

\subsection{``Pickiness''}
\label{ap:betas}
Under the \qdist{} formalism, we cannot determine how BHs in binaries choose their companions, though we can gain some insight from their marginal mass ratio distribution. 
\qdist{} parametrizes the mass ratio distribution as a power law with spectral index $\beta$, where $\beta>0$ corresponds to mass ratio distributions with more support for similar-mass binaries.
\cite{abbott_population_2023} find $\beta = 1.1^{+1.7}_{-1.3}$, which similarly indicates a preference for equal-mass binaries.
Note that the value for $\beta$ under \qdist{} is noticeably different than the value found for $\beta_q$ under the pairing function models because the two parameters cause different behaviors in the 2D mass distribution within their respective models.
A low value for $\beta$ does not imply that BBHs are not picky, just that the marginal mass ratio distribution rises slowly.
As shown in Appendix~\ref{ap:model comparison}, the pairing function models and \qdist{} model all produce near-identical marginal mass ratio distributions, despite different values for $\beta_q$ and $\beta$.
Low values for $\beta$ are partially due to the fact that the low-$q$ end of the marginal mass ratio distribution will always be suppressed because of the existence of a minimum BH mass.
This minimum mass makes it impossible to get extreme mass ratios unless $m_1$ is very large, and the $m_1$ distribution has very little support for $m_1\gtrapprox60\Msun$.
Therefore, $\beta$ does not need to be large in order for the marginal mass ratio distribution to strongly disfavor unequal-mass binaries.
In fact, the existence of a minimum mass plus a tapering at high $m_1$ means that even when $\beta <0$, the marginal mass ratio distribution rises towards $q=1$.
\comment{We could add a plot of this if its helpful.}

\subsection{Mimicking \qdist{} with a pairing function model}
\label{ap:all_same_lam2_0}
We show that the \lamTwoeqZero{} model approximates the morphology of the \textsc{LVK 2023} model, which uses the \qdist{} formalism.

The right two panels of Figure~\ref{fig:lam2_0 vs qdist} show the two-dimensional PPDs for the \textsc{LVK 2023} and \lamTwoeqZero{} models.
Both exhibit vertical bands and no horizontal bands, and have a similar-magnitude drop in merger rate moving away from the diagonal.

The leftmost panel of Figure~\ref{fig:lam2_0 vs qdist} shows the conditional $m_2$ distribution for both models.
While the slopes of the two models differ slightly, a peak in $m_2$ appears in neither.
We therefore find \lamTwoeqZero{} to be an appropriate proxy for the behavior of the \textsc{LVK 2023} model for the purposes of this work.
A future measurement ruling out $\lambda_2=0$ with high confidence would therefore serve as evidence for the pairing function models over \textsc{LVK 2023}-like models.

\section{General Population Model}
\label{ap:most general model}
\subsection{Base model}
\label{sec:mass model}
We model the two-dimensional mass distribution by constructing separate one-dimensional distributions for the primary and secondary masses and combining draws from these two distributions according to a pairing function, as in Equation~\ref{eq:pairing function}.
\citet{fishbach_picky_2020} find the pairing function is most informative when parameterized by the mass ratio of the binary, so we adopt a pairing function described by a power law in mass ratio.
    
Because our aim is to learn whether the primary and secondary masses are consistent with being drawn from the same distribution (up to a pairing function and subject to the constraint that $m_1\geq m_2$), we describe the primary and secondary mass distributions separately, but according to the same functional form.
We model each of the 1-dimensional mass distributions as a mixture model between a smoothed power law component and a Gaussian component $G$ in order to make direct comparisons to the \textsc{Power Law + Peak} model used by the LIGO-Virgo-KAGRA's (LVK) population analysis to describe the distribution of primary masses \cite{talbot_measuring_2018,abbott_population_2023}.
Explicitly,
\begin{equation}
\begin{aligned}
    p(m_1|\Lambda_1) \propto \bigg[&
(1-\lambda_1)\Theta(m_1> m_{\min,1})\Theta(m_1<m_{\max,1}) \left( \frac{\alpha_1 + 1}{m_{\max,1}^{\alpha_1 +1} - m_{\min,1}^{\alpha_1+1}} \right) m_1^{-\alpha_1} +
\lambda_1 G(m_1|\mu_1,\sigma_1)
\bigg]\\
& \times S(m_1|m_{\min,1}, \delta_1) \\
p(m_2|\Lambda_2) \propto \bigg[&
(1-\lambda_2)\Theta(m_2> m_{\min,2})\Theta(m_2<m_{\max,2}) \left( \frac{\alpha_2 + 1}{m_{\max,2}^{\alpha_2 +1} - m_{\min,2}^{\alpha_2+1}} \right) m_2^{-\alpha_2} +
\lambda_2 G(m_2|\mu_2,\sigma_2)
\bigg]\\
& \times S(m_2|m_{\min,2}, \delta_2)
.
\end{aligned}
\label{eq:1d mass}
\end{equation}

Here, $G(m_{\{1,2\}}|\mu_{\{1,2\}}, \sigma_{\{1,2\}})$ is a normalized Gaussian distribution with mean $\mu_{\{1,2\}}$ and width $\sigma_{\{1,2\}}$.
The parameter $\lambda_{\{1,2\}}$ is a mixing fraction determining the relative prevalence of mergers the power law and Gaussian components, and $S(m_{\{1,2\}}, m_{\min,{\{1,2\}}} \delta_{\{1,2\}})$ is a smoothing function which rises from 0 to 1 over the interval $(m_{\min,{\{1,2\}}}, m_{\min,{\{1,2\}}}+\delta_{\{1,2\}})$.
$\Lambda_1$ is then the set of hyper-parameters $\{ m_{\min,1}, m_{\max,1}, \alpha_1, \lambda_1, \mu_1, \sigma_1, \delta_1 \}$ and $\Lambda_2 = \{m_{2 \min} m_{\max,2}, \alpha_2, \lambda_2, \mu_2, \sigma_2, \delta_2\}$.

In all models considered in this work, the redshift distribution is modeled as a power law with spectral index $\kappa$ \citep{fishbach_does_2018} such that
\begin{equation}
    p(z) \propto \frac{dV_c}{dz} \left(\frac{1}{1+z}\right) (1+z)^\kappa.
\end{equation}
We use the \textsc{Default} spin model from \citet{abbott_population_2021,abbott_population_2023} to describe the spin magnitudes and tilts of each component.
These are the same redshift and spin distributions used with the \textsc{Power Law + Peak} mass model in the analysis presented in \citet{abbott_population_2023}.

A fit to the model described above is provided in Appendix~\ref{ap:most general model}, though we focus on specific variations nested within this more general model for the remainder of this work.

\subsection{Fit to general base model}
We present results of a fit to the most general form of the pairing function model described in Section~\ref{sec:mass model} (Equation~\ref{eq:1d mass}).
Here, we do not fix any parameters to be equal between $\Lambda_1$ and $\Lambda_2$, and instead infer them separately.

\begin{figure*}
    \includegraphics[width=\textwidth]{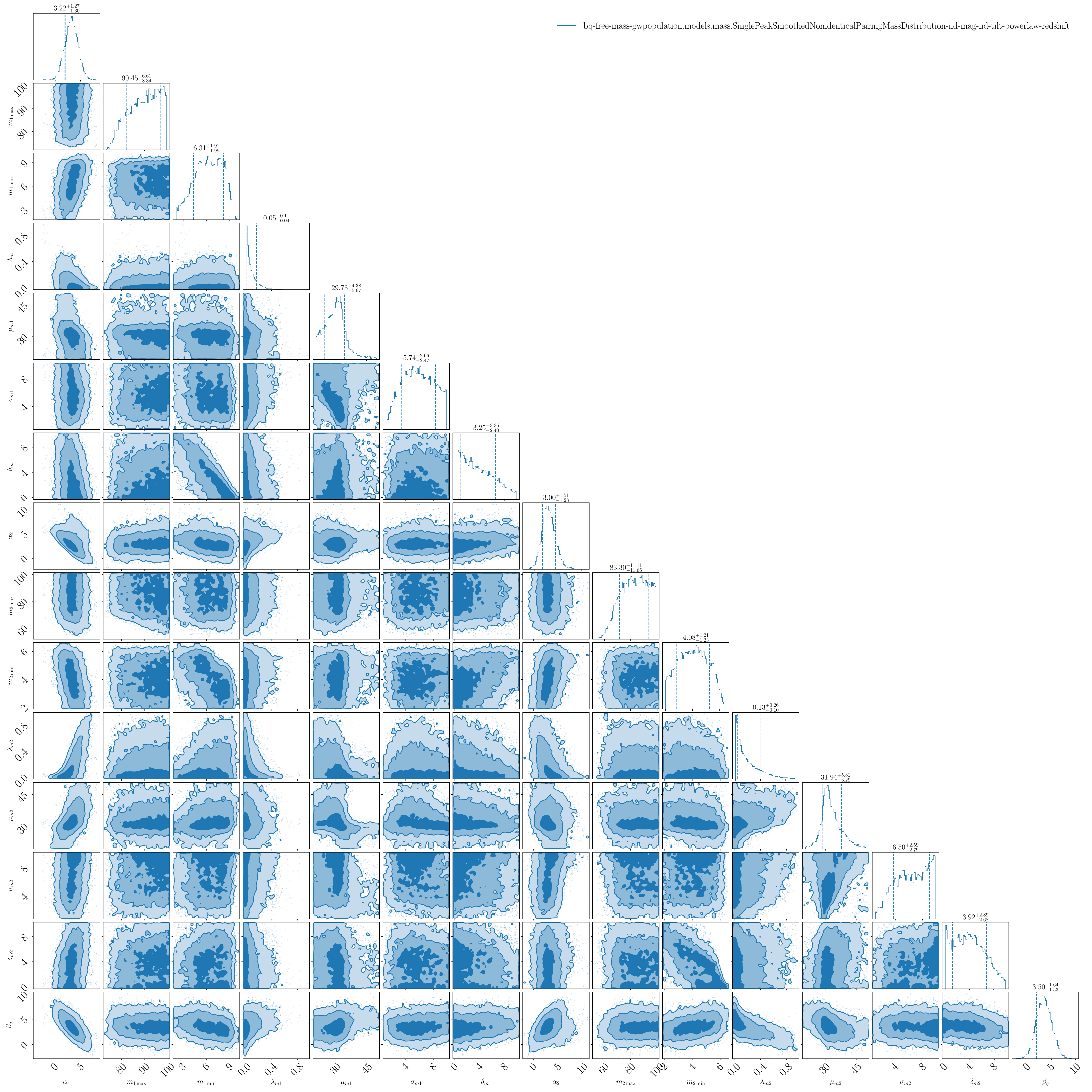}
    \caption{Corner plot of all mass-related hyper-posteriors in the most general form of the base model.
    Power law spectral indecies governing the slope of the primary and secondary mass distributions are degenerate with one another and with the pairing function power law, so only two of the three parameters can be meaningfully constrained at a time.
    All other parameters that perform the same function in $p_1(m_1)$ and $p_2(m_2)$ are consistent between the two underlying distributions.
    }
    \label{fig:base model corner}
\end{figure*}

Figure~\ref{fig:base model corner} shows the posterior of all mass-related hyper-parameters within this model in the form of a corner plot.
Because of the large number of free parameters, several are not well-constrained.
Nonetheless, all parameters describing $p_1(m_1)$ are consistent with those describing $p_2(m_2)$.
Additionally, strong correlations exist between $\alpha_1,\alpha_2$, and $\beta_q$, making it difficult to meaningfully constrain all three parameters.
We therefore set $\alpha_1=\alpha_2$ in the \diff{} model.

\begin{figure}
    \centering
    \includegraphics[width=0.7\textwidth]{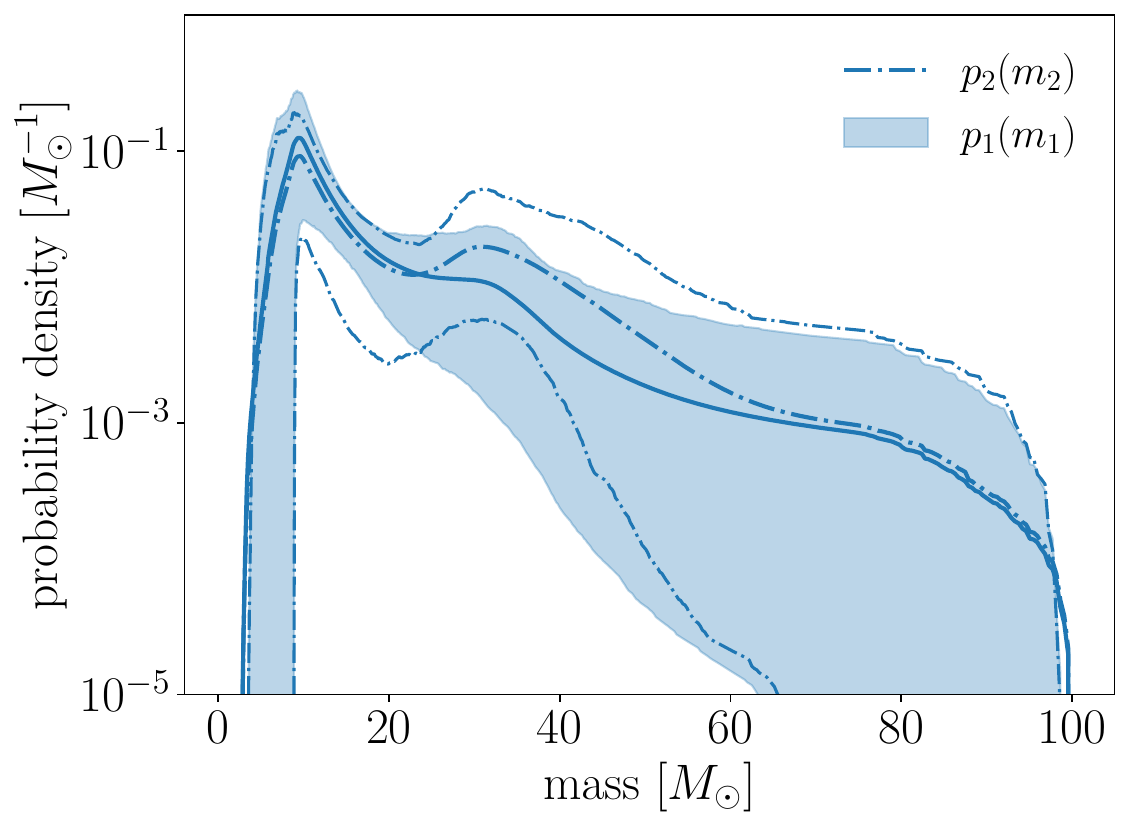}
    \caption{Underlying distributions (i.e. before a pairing function is applied) of the primary (\emph{shaded band}) and secondary (\emph{dot-dashed lines}) masses under the most general form of the base model.
    Despite fitting all parameters separately between the two distributions, they appear consistent with one another. 
    However, both are relatively poorly constrained.
    Interestingly, the support for a peak in $p_1(m_1)$ lessens in this more generalized model, while the support for a peak in $p_2(m_2)$ remains the same as in \diff.
    }
    \label{fig:base model bucket}
\end{figure}

Figure~\ref{fig:base model bucket} shows the underlying distributions, $p_1(m_1)$ and $p_2(m_2)$ inferred by the most general version of the base model.
As expected, the constraints on these distributions weaken relative to that of the \diff{} model.
However it is still clear that $p_1(m_1)$ and $p_2(m_2)$ are consistent with one another and that it is possibile that $p_2(m_2)$ has a larger peak than $p_1(m_1)$ does.
Notably, there now seems to be very little evidence for a peak in $p_1(m_1)$, whereas a clear peak still exists in $p_2(m_2)$.

These results are consistent with what has been presented using the \diff{} model in the main body of this work.

\subsection{Model comparison}
\label{ap:model comparison}
In this Section, we compare the various models considered in this work, including the \qdist{} model from \textsc{LVK 2023}.
Table~\ref{tab:models} lists the hyper-prior choices made to construct each model, along with descriptions of each hyper-parameter.

\begin{figure}
    \centering
    \includegraphics[width=0.7\textwidth]{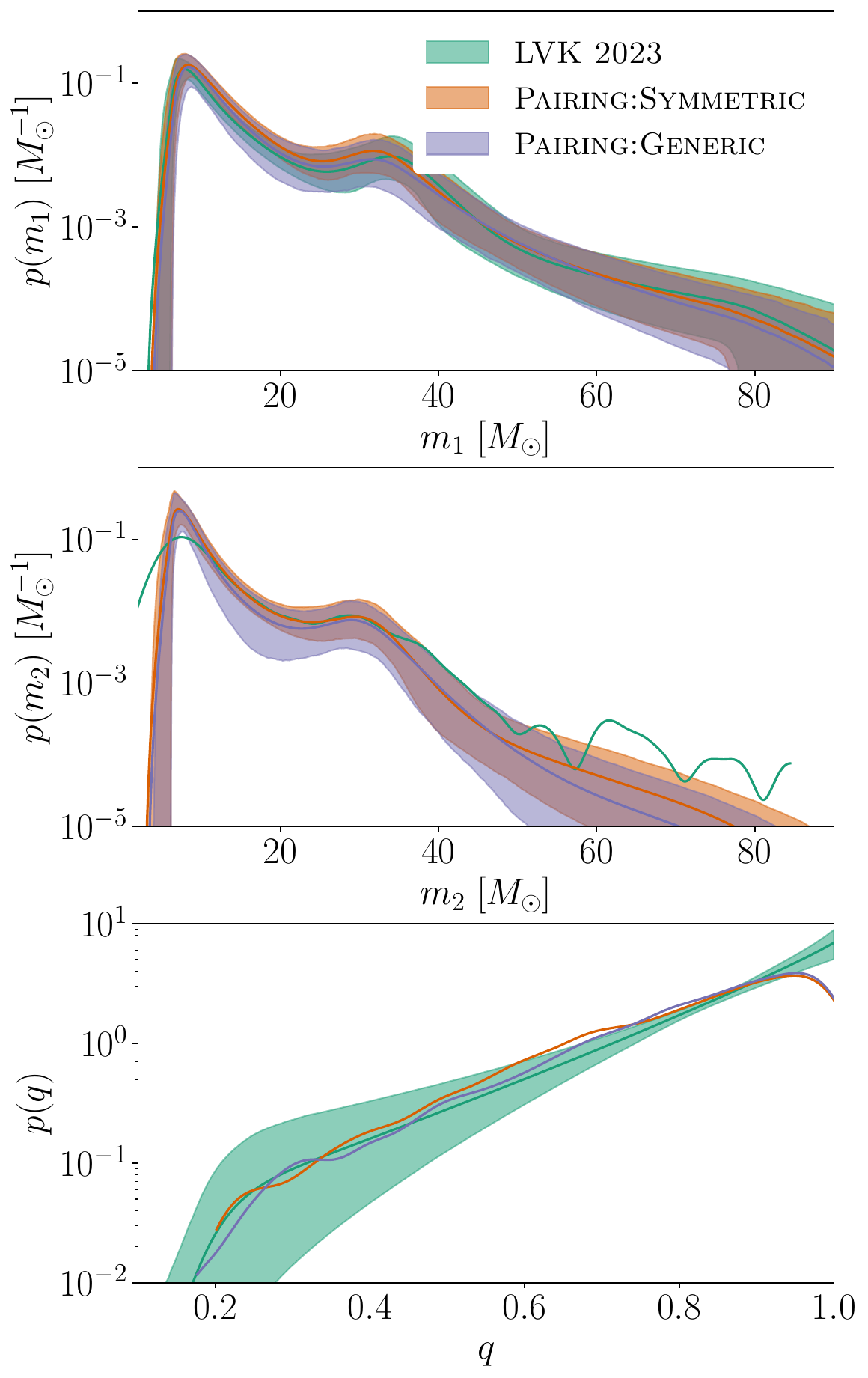}
    \caption{Marginal PPDs of primary mass (\emph{top}), secondary mass (\emph{middle}) and mass ratio (\emph{bottom}) for \qdist{} (\emph{green}), \default{} (\emph{orange}) and \diff{} (\emph{purple}). Solid lines are the mean value and shaded regions represent the 90\% credible interval. These are obtained by integrating the 2D PPDs shown in Figure~\ref{fig:2d ppds} along each dimension in turn.
    Since the \qdist{} model is parameterized in terms of $m_1$ and $q$, we reconstruct its marginal $m_2$ distribution by sampling from the joint distribution and creating a kernel density estimate, causing some artificial ``wiggles’’ in these plots. 
    The same is true for the marginal $q$ distributions of the pairing function models.
    Despite the 2D PPDs appearing different between \qdist{} and the other two models, the marginal PPDs all appear similar.
    This is because features in the primary mass distribution induce features in the marginal secondary mass distribution if equal-mass binaries are preferred. 
    Therefore, marginal PPDs are not sensitive to the features in the 2D mass distribution in which we are interested.
}
    \label{fig:marginal ppds}
\end{figure}

\begin{figure}
    \centering
    \subfloat{
         \centering
         \includegraphics[width=0.32\textwidth]{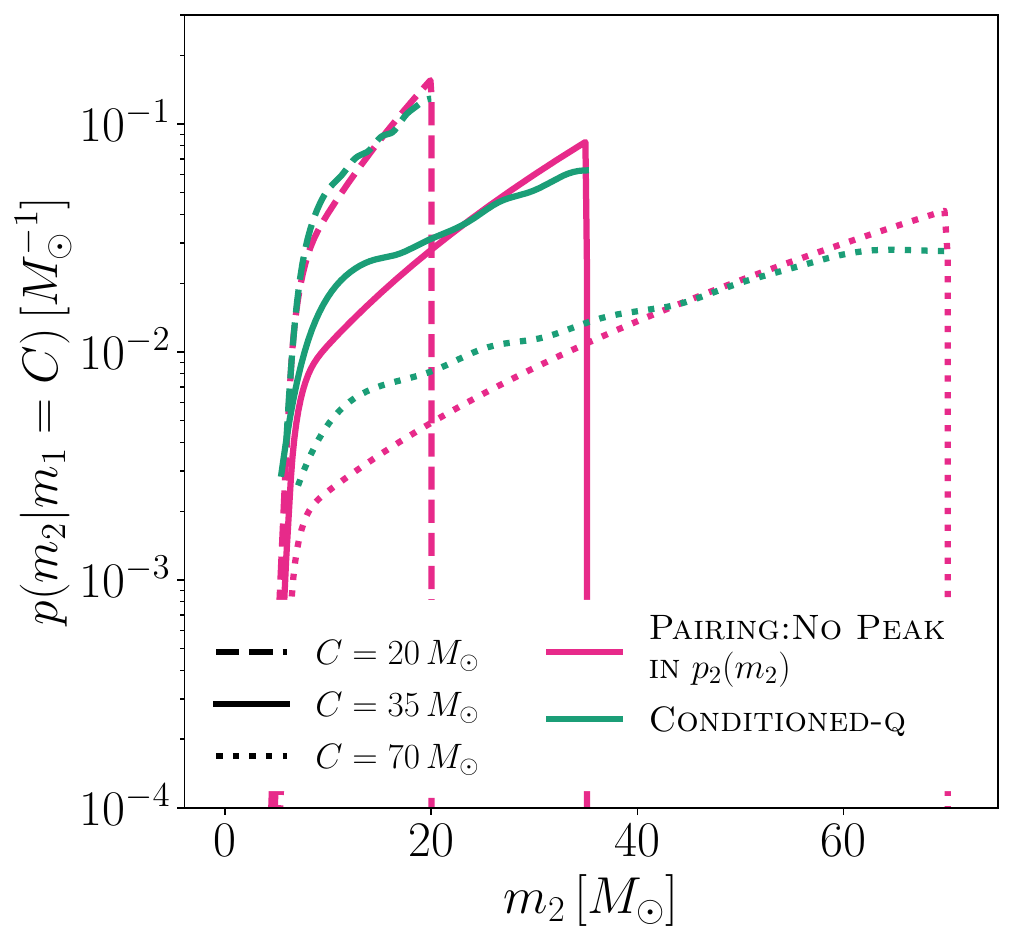}
     }
     \hfill
    \subfloat{
         \centering
         \includegraphics[width=0.65\textwidth]{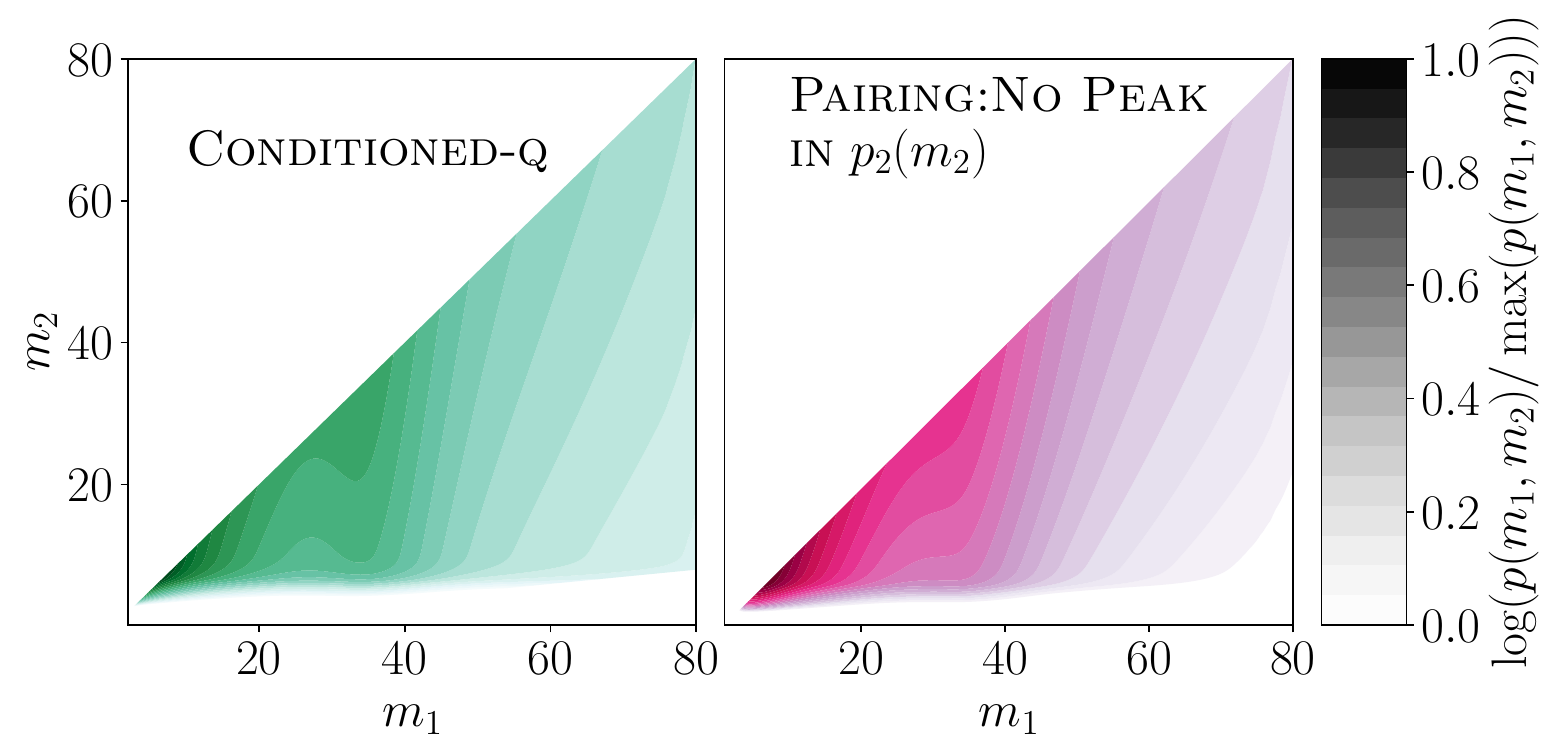}
     }     
     \caption{Two-dimensional  (\emph{right}) and conditional $m_2$ (\emph{left}) PPDs for the \qdist{} and \lamTwoeqZero{} models.}
     \label{fig:lam2_0 vs qdist}
\end{figure}

Figure~\ref{fig:marginal ppds} shows the $m_1$, $m_2$, and $q$ PPDs, marginalized over all other parameters.
These show a high level agreement between the different models, despite their two-dimensional PPDs differing in Figure~\ref{fig:2d ppds}.
Notably, the marginal secondary mass distribution exhibits a peak even for the \qdist{} model.
As discussed in Appendix~\ref{sec:pairing-vs-pq-pedagogy}, this is caused by the preference for equal masses: if $m_1$ is in the $\sim35\Msun$ peak, a preference for $m_2\approx m_1$ will cause $m_2$ to also preferentially be in that region, causing more probability density for $m_2\sim35\Msun$ than elsewhere.
Marginal PPDs are therefore not very sensitive to whether $m_2$ prefers its own features independently of the existence of a preference for equal mass binaries.
Instead, we examine the underlying component mass distributions explicitly modeled by \default{} and \diff, as well as the values of informative hyper-parameters in these models to determine whether $p_2(m_2)$ differs from $p_1(m_1)$.
For comparisons to \qdist, we plot secondary mass distributions conditioned on various values of $m_1$.

\begin{table}
    \centering
    \begin{tabular}{p{0.5cm} p{5cm} c | c | c | c | c}
        \hline \hline
         \multicolumn{1}{c}{\multirow{3}{*}{\textbf{Parameter}}} & \multicolumn{1}{c}{\multirow{3}{*}{\textbf{Description}}} & \multicolumn{4}{c}{\textbf{Prior}} \\
           & &  & \multirow{2}{*}{Base model} & \textsc{Pairing:} & \textsc{Pairing:} & \multirow{2}{*}{\qdist} \\
           & &  & & \textsc{Generic} & \textsc{Symmetric} & \\
        \hline \hline
        \multirow{2}{*}{$\beta_q$} & Spectral index for the power law of the pairing function. & & \multicolumn{3}{c|}{\multirow{2}{*}{$\mathrm{U}(-4, 12)$}}&\multirow{2}{*}{--} \\
        \hline
        \multirow{2}{*}{$\beta$} & Spectral index for the power law of the mass ratio distribution. & & \multicolumn{3}{c|}{\multirow{2}{*}{--}}&\multirow{2}{*}{$\mathrm{U}(-4, 12)$} \\
         \hline
        \multirow{2}{*}{$\alpha_1$} & Spectral index for the power law of the primary mass distribution & & \multicolumn{4}{c}{\multirow{2}{*}{$\mathrm{U}(-4, 12)$}} \\
        \hline
        \multirow{2}{*}{$\alpha_2$} & Spectral index for the power law of the secondary mass distribution & & \multirow{2}{*}{$\mathrm{U}(-4, 12)$}&\multicolumn{2}{c|}{\multirow{2}{*}{$\alpha_2=\alpha_1$}} & \multirow{2}{*}{--}\\
        \hline
        \multirow{2}{*}{$m_{\min,1}$} & Minimum mass of the primary mass distribution. &  & \multicolumn{4}{c}{\multirow{2}{*}{U($2\, M_{\odot}$, $10\, M_{\odot}$)}}\\
        \hline
        \multirow{2}{*}{$m_{\min,2}$} & Minimum mass of the secondary mass distribution. &  & \multirow{2}{*}{U($2\, M_{\odot}$, $10\, M_{\odot}$)}& \multicolumn{2}{c|}{\multirow{2}{*}{$m_{\min,2}=m_{\min,1}$}} &  \multirow{2}{*}{--} \\
        \hline
        \multirow{2}{*}{$m_{\max,1}$} & Maximum mass of the primary mass distribution. &  & \multicolumn{4}{c}{\multirow{2}{*}{U($30\, M_{\odot}$, $100\, M_{\odot}$)}}\\
        \hline
        \multirow{2}{*}{$m_{\max,2}$} & Maximum mass of the secondary mass distribution. &  & \multirow{2}{*}{U($30\, M_{\odot}$, $100\, M_{\odot}$)}& \multicolumn{2}{c|}{\multirow{2}{*}{$m_{\max,2}=m_{\max,1}$}} &  \multirow{2}{*}{--} \\
        \hline
        \multirow{2}{*}{$\delta_{1}$} & Range of tapering at the low end of the primary mass distribution. &  & \multicolumn{4}{c}{\multirow{2}{*}{U($0\, M_{\odot}$, $10\, M_{\odot}$)}}\\
        \hline
        \multirow{2}{*}{$\delta_{2}$} & Range of tapering at the low end of the secondary mass distribution. &  & \multirow{2}{*}{U($0\, M_{\odot}$, $10\, M_{\odot}$)}& \multicolumn{2}{c|}{\multirow{2}{*}{$\delta_2=\delta_1$}} &  \multirow{2}{*}{--} \\
        \hline    
        \multirow{2}{*}{$\lambda_1$} & Fraction of systems with primary mass in the Gaussian component. &  & \multicolumn{4}{c}{\multirow{2}{*}{U(0, 1)}} \\
        \hline
        \multirow{2}{*}{$\lambda_2$} & Fraction of systems with secondary mass in the Gaussian component. &  & \multicolumn{2}{c|}{\multirow{2}{*}{U(0,1)}}& \multirow{2}{*}{$\lambda_2=\lambda_1$} &  \multirow{2}{*}{--} \\
        \hline
        \multirow{2}{*}{$\mu_1$} & Mean of the Gaussian component in the primary mass distribution. &  & \multicolumn{4}{c}{\multirow{2}{*}{U($20\, M_{\odot}$, $50\, M_{\odot}$)}} \\
        \hline
        \multirow{2}{*}{$\mu_2$} & Mean of the Gaussian component in the secondary mass distribution. &  & \multicolumn{2}{c|}{\multirow{2}{*}{U($20\, M_{\odot}$, $50\, M_{\odot}$)}}& \multirow{2}{*}{$\mu_2=\mu_1$} &  \multirow{2}{*}{--} \\
        \hline
        \multirow{2}{*}{$\sigma_1$} & Width of the Gaussian component in the primary mass distribution. &  & \multicolumn{4}{c}{\multirow{2}{*}{U($1\, M_{\odot}$, $10\, M_{\odot}$)}} \\
        \hline
        \multirow{2}{*}{$\sigma_2$} & Width of the Gaussian component in the secondary mass distribution. &  & \multicolumn{2}{c|}{\multirow{2}{*}{U($1\, M_{\odot}$, $10\, M_{\odot}$)}}& \multirow{2}{*}{$\sigma_2=\sigma_1$} &  \multirow{2}{*}{--} \\
        \hline
    \end{tabular}
    \caption{
    Hyperparameters of our mass model and hyper-priors corresponding to specific model variations.
    For comparison, we also show the hyper-priors for the \textsc{LVK 2023} model, which uses the \qdist{} framework.
    We denote the uniform distribution between $x$ and $y$ as $\mathrm{U}(x, y)$, list specific values that are fixed in some priors, and denote when hyper-parameters are irrelevant to a specific nested model with ``--''.
    }
    \label{tab:models}
\end{table}

\section{Statistical Framework}
\label{ap:HBA}
In this Appendix, we describe the hierarchical Bayesian analysis used to fit the population models described in Appendix~\ref{ap:most general model} and Section\ref{sec:methods} to BBHs in GWTC-3  \citep{abbott_gwtc-3_2021, abbott_open_2023}.

The posterior on the population hyper-parameters, assuming a prior on the overall rate of mergers $p(\mathcal{R}) \sim 1/\mathcal{R}$ and marginalizing, is
\begin{equation}
    p(\Lambda |\{D_j\}) = p(\Lambda) \prod_j^N \frac{p(D_j|\Lambda)}{\mathcal{E}(\Lambda)},
    \label{eq:hyper-posterior}
\end{equation}
where
\begin{equation}\label{eq:marginal likelihood}
    p(D_j|\Lambda) = \int dm_1 dm_2 ds_1 ds_2 dz \, p(z) p(s_1,s_2) p(m_1, m_2|\Lambda) p(D_j|m_1, m_2, s_1, s_2, z)
\end{equation}
is the marginal likelihood for the $j^{\rm th}$ event,
\begin{equation}\label{eq:selection effects}
    \mathcal{E}(\Lambda) = \int dm_1 dm_2 ds_1 ds_2 dz \, p(z) p(s_1,s_2) p(m_1, m_2|\Lambda) P(\mathrm{det}|m_1, m_2, s_1, s_2, z)
\end{equation}
is the fraction of detectable events in a population described by $\Lambda$, and $P(\mathrm{det}|m_1, m_2, s_1, s_2, z)$ is the probability that any individual event with parameters $m_1$, $m_2$, $s_1$, $s_2$, and $z$ would be detected, averaged over the duration of the experiment.

In practice, the high-dimensional integrals in Eqs.~\ref{eq:marginal likelihood} and~\ref{eq:selection effects} are approximated via importance sampling \citep[see][ for a detailed explanation of this process]{essick_precision_2022,essick_estimating_2021}.
Given a set of $N_j$ event-level samples drawn from the posterior for the $j^\mathrm{th}$ event that used a reference prior $p_\mathrm{ref}(m_1, m_2, s_1, s_2, z)$, we approximate
\begin{equation}\label{eq:approx marginal likelihood}
    \frac{p(D_j|\Lambda)}{p_\mathrm{ref}(D_j)} \approx \frac{1}{N_j}\sum\limits_k^{N_j} \frac{p(m_1^{(k)}, m_2^{(k)}, s_1^{(k)}, s_2^{(k)}, z^{(k)}|\Lambda)}{p_\mathrm{ref}(m_1^{(k)}, m_2^{(k)}, s_1^{(k)}, s_2^{(k)}, z^{(k)})}
\end{equation}
where $p_\mathrm{ref}(D_j)$, the marginal likelihood for $D_j$ under the reference prior, does not depend on $\Lambda$ and therefore need not be included when determining the population fit.
Similarly, by simulating a large set of $M$ signals drawn from an injected population $p_\mathrm{draw}$, we can approximate Eq.~\ref{eq:selection effects} with a sum over the subset of $m$ detected signals
\begin{equation}\label{eq:approx selection effects}
    \mathcal{E}(\Lambda) \approx \frac{1}{M}\sum\limits_k^m \frac{p(m_1^{(k)}, m_2^{(k)}, s_1^{(k)}, s_2^{(k)}, z^{(k)}|\Lambda)}{p_\mathrm{draw}(m_1^{(k)}, m_2^{(k)}, s_1^{(k)}, s_2^{(k)}, z^{(k)})}
\end{equation}
We obtain the detectable set of $m$ signals by injecting waveforms of these signals into the measured detector strain and running the search pipelines \citep{usman_pycbc_2016,aubin_mbta_2021,drago_coherent_2021,cannon_gstlal_2021} used by the LVK to detect events to obtain a false alarm rate (FAR) for each injected signal.
This process was performed by the LVK for its population analysis \citep{abbott_population_2023} and we use the resulting publicly-released data product in this work \citep{collaboration_gwtc-3_2021}.
We then consider an event detected if it has a FAR of less than 1/year in at least one pipeline, and apply this threshold both to injected signals and the real GW events in GWTC-3.

We sample from the posterior distribution in Eq.~\ref{eq:hyper-posterior} using the approximations in Eqs.~\ref{eq:approx marginal likelihood} and~\ref{eq:approx selection effects} to determine the shape of the mass distribution using \texttt{gwpopulation} \citep{talbot_parallelized_2019} with the nested sampling algorithm \texttt{dynesty} \citep{speagle_dynesty_2020}. 
Furthermore, where needed, we estimate Bayes factors via Savage-Dickey Density Ratios \citep{dickey_weighted_1970,wagenmakers_bayesian_2010} using the hyper-posteriors and the hyper-priors described in Section~\ref{sec:mass model}.

\nocite{olsen_two_1998}
\bibliography{references}{}
\bibliographystyle{aasjournal}



\end{document}